\documentclass[aps,prb,reprint,superscriptaddress]{revtex4-2}

\usepackage{hyperref}
\usepackage{graphicx}
\usepackage{array}        
\usepackage{amsmath}      
\usepackage{amssymb}      
\usepackage{bm}           
\usepackage{color}        
\usepackage{enumitem}     
\usepackage{comment}      
\usepackage{braket}       
\usepackage{longtable}    
\usepackage{lipsum}       
\usepackage{mathrsfs}     

\newcommand{\pt}{$\mathcal{PT}$}

\begin{document}

\title{Morse theory study on the evolution of nodal lines in $\mathcal{PT}$-symmetric nodal-line semimetals}
\author{Manabu Takeichi}
\affiliation{Department of Physics, Tokyo Institute of Technology, 2-12-1 Ookayama, Meguro-ku, Tokyo 152-8551, Japan}
\author{Ryo Furuta}
\affiliation{Department of Physics, Tokyo Institute of Technology, 2-12-1 Ookayama, Meguro-ku, Tokyo 152-8551, Japan}
\author{Shuichi Murakami}
\affiliation{Department of Physics, Tokyo Institute of Technology, 2-12-1 Ookayama, Meguro-ku, Tokyo 152-8551, Japan}
\date{\today}

\begin{abstract}
A nodal-line semimetal is a  topological gapless phase containing one-dimensional degeneracies called nodal lines. The nodal lines are deformed by a continuous change of the system such as pressure and they can even change their topology, but it is  not systematically understood what kind of changes of topology of nodal lines are possible. In this paper, we classify the events of topology change of nodal lines by the Morse theory and reveal that only three types of topology changes of nodal lines, i.e.,  creation, reconnection, and annihilation, are possible in the spinless nodal-line semimetal protected by inversion and time-reversal symmetries. They are characterized by an index having the values 0, 1, and 2 for the above three types in  the Morse theory. Moreover, we extend our theory to systems with  rotational symmetries and mirror symmetry, and disclose the possible events of topology change of nodal lines under each symmetry.
\end{abstract}

\maketitle

\section{introduction}
Topological phases of matter have been attracting much attention in condensed matter physics. The topological phases are classified into two cases: topological insulators \cite{RevModPhys.83.1057,RevModPhys.82.3045,PhysRevB.76.045302,PhysRevLett.95.146802,PhysRevLett.95.226801,PhysRevLett.98.106803,doi:10.1126/science.1133734}  and topological semimetals \cite{RevModPhys.90.015001,Fang_2016,doi:10.1146/annurev-matsci-070218-010049,RevModPhys.93.025002}. The topological semimetals can hold either zero-dimensional (0D), one-dimensional (1D), or two-dimensional (2D) degeneracies in $\bm k$ space between a conduction band and a valence band in three-dimensional (3D) materials. In the momentum space, a topological 0D degeneracy is  Dirac points \cite{PhysRevB.85.195320,PhysRevLett.108.140405,Yang2014} and Weyl points \cite{PhysRevB.83.205101,PhysRevB.85.155118,PhysRevLett.107.127205,doi:10.1126/science.aaf5037,Fang2016,PhysRevLett.119.206401,Shuichi_Murakami_2007,Soluyanov2015,PhysRevX.5.011029,PhysRevX.5.031013,pnas.1514581113,Chang2018}, a topological 1D degeneracy is nodal lines \cite{Fang_2016,ParkGaoZhangOh+2022+2779+2801}, and a topological 2D degeneracy is nodal surfaces \cite{doi:10.1126/sciadv.aau6459,doi:10.1126/sciadv.aav2360,PhysRevB.97.115125,PhysRevB.96.155105,PhysRevB.93.085427,PhysRevB.105.245152,Yang2019}. These gapless states are robust against perturbations because of symmetry or topological reasons. The topological semimetals possessing the nodal lines are called nodal-line semimetals, and the nodal lines have several varieties depending on the relative positions of nodal lines: nodal rings \cite{PhysRevB.92.081201,PhysRevB.84.235126,PhysRevB.93.205132,Schoop2016,Chen2015,PhysRevLett.115.036806,PhysRevLett.115.026403}, nodal chains \cite{Bzdusek2016,PhysRevLett.119.036401,Yan2018}, nodal links \cite{PhysRevB.96.041102,PhysRevB.96.041103,PhysRevB.97.155140,doi:10.1021/acsphotonics.1c00876,PhysRevLett.125.033901,PhysRevLett.128.246601}, and so on. The nodal ring is a loop of the nodal line, the nodal chain has touching points of two nodal lines, and the nodal link forms a link between two nodal lines.

The nodal line is protected by crystal symmetry, such as mirror symmetry or a combination of inversion ($\mathcal{P}$) symmetry and time-reversal  ($\mathcal{T}$) symmetry \cite{Fang_2016,doi:10.1146/annurev-matsci-070218-010049}. The nodal lines are confined on the mirror planes in the former case, and there are no constraints for the positions of nodal lines in the latter case. Meanwhile, the nodal lines are characterized by a quantized value of the Berry phase \cite{berry1984quantal,RevModPhys.82.1959,vanderbilt_2018} in the latter. When the spin-orbit coupling (SOC) is negligible in systems considered, the Berry phase on any closed path is quantized to be 0 or $\pi$ modulo $2\pi$ under \pt\ symmetry. The nodal line with \pt\ symmetry has a $\pi$ Berry phase when the closed path links with the nodal line.

In this paper, we focus on the nodal lines protected by the $\pi$ Berry phase.
Under a continuous change of the system, the shapes of the nodal lines with the $\pi$ Berry phase are deformed  as long as the system keeps the \pt\ symmetry.  In addition to the deformations of the shapes of nodal lines, the nodal lines may change their connectivity, i.e., their topology. For example, through a continuous change of the system, two nodal lines may merge into one and vice versa.

In this paper, we show that the change of topology of nodal lines are classified in terms of the Morse theory and reveal that there are only three cases for the change of topology, i.e., creation, reconnection, and annihilation. We introduce the notion of the index in the Morse theory. Moreover, we classify the evolutions of nodal lines in systems with mirror or rotational symmetry.
In these cases with additional crystallographic symmetry, the Morse theory cannot be directly applied and the index is not defined. Here, we find that in such cases the coefficient functions in the Hamiltonian is always ``factorized," and after the factorization one can apply the Morse theory to define the index and to classify the events of topology changes of nodal lines. Through this study, we exhaust all the possible events of topology changes of nodal lines.
It also means that events of topology changes other than these listed in this paper do not occur. For example, a direct transition from two nodal lines to a nodal link cannot occur when no crystallographic symmetry is assumed.

This paper is organized as follows. In Sec. \ref{sec:NL_with_pi_berry_pahse}, we show an example of the evolution of the nodal line in a previous study and the limitation of the model. In Sec. \ref{sec:evolution_no_sym}, we reveal relationships between an index and a possible evolution of nodal line under \pt\ symmetry and classify the changes with indices. In Sec. \ref{sec:evolution_with_sym}, we show  evolutions of nodal lines  with an additional rotational or mirror symmetry. We summarize the paper in Sec. \ref{sec:summary}.

\section{Nodal lines with $\pi$ Berry phase and their evolutions \label{sec:NL_with_pi_berry_pahse}}

\subsection{Nodal lines with $\pi$ Berry phase}
We study nodal lines in 3D spinless systems protected by the quantized $\pi$ Berry phase. For this purpose, we need to consider one conduction and one valence bands, and a two-band Hamiltonian is written as
\begin{align}
	\mathcal{H}(\bm k)=a_0(\bm k)\sigma_0+\bm a(\bm k)\cdot\bm\sigma,
\end{align}
where  $\bm k=(k_x,k_y,k_z)$, $a_0(\bm k)$ and $\bm a(\bm k)=(a_x(\bm k), a_y(\bm k), a_z(\bm k))$ are real functions, $\sigma_0$ is the $2\times 2$ identity matrix, and $\bm\sigma=(\sigma_x,\sigma_y,\sigma_z)$ are the Pauli matrices. We put $a_0(\bm k)=0$ for simplicity because it does not affect the nodal lines.
In the absence of SOC,  \pt\  symmetry ensures that the Hamiltonian is real ($\mathcal{H}(\bm k)=\mathcal{H}^{*}(\bm k)$) i.e., $a_y(\bm k)=0$ under an appropriate gauge choice. The energy spectra are $E_{\pm}(\bm k)=\pm\sqrt{a_x^2(\bm k)+a_z^2(\bm k)}$, and the positions of the nodal lines in $\bm k$ space are obtained by solving $a_x(\bm k)=a_z(\bm k)=0$.
The Berry phase along a loop in the $\bm k$ space in a system with \pt\ symmetry without SOC is quantized to 0 or $\pi$ modulo $2\pi$, and the nodal line is protected by the $\pi$ Berry phase. 

\subsection{Change of topology of the nodal lines\label{sec:previous-nodal-link}}
In the previous study \cite{PhysRevB.96.041103},  a two-band model for a nodal link and a nodal chain is proposed, and the Hamiltonian $\mathcal{H}^{(\mathrm{A})} (\bm k)$ is given as
\begin{align}
	\mathcal{H}^{(\mathrm{A})} (\bm k)= &a^{(\mathrm{A})}_x(\bm k)\sigma_x+a^{(\mathrm{A})}_z(\bm k)\sigma_z \label{eq:H_A} \\
	a^{(\mathrm{A})}_x(\bm k) = &2\sin k_x\sin k_z + 2f(\bm k)\sin k_y, \label{eq:a_Ax} \\
	a^{(\mathrm{A})}_z(\bm k) = &\sin^2 k_x+\sin^2k_y-\sin^2k_z - f^2(\bm k), \label{eq:a_Az}
\end{align}
where $f(\bm k)=\sum_{i=x,y,z}\cos k_i-m$ and $m$ is a real parameter.
This two-band model exhibits two nodal rings for $m>3$, a nodal chain with a touching point $\bm k=0$ for $m=3$, and a nodal link for $m<3$.  Namely, the nodal chain is an intermediate state between the nodal rings and the nodal link. However, we will show that this kind of a direct change between the nodal ring and the nodal link via the nodal chain does not occur in general, and this change in the model is permitted due to a special feature of the model, which is explained further below.

The nodal lines are regarded as intersections between two 2D closed surfaces  $a^{(\mathrm{A})}_x(\bm k)=0$ and $a^{(\mathrm{A})}_z(\bm k)=0$ in the momentum space. By using this, we can calculate the tangential vector of the nodal line. The normal vector $\bm n^{(x)}$ of the closed surface $a^{(\mathrm{A})}_x(\bm k)=0$  is parallel to $\nabla_{\bm k}a^{(\mathrm{A})}_x=(\frac{\partial a^{(\mathrm{A})}_x}{\partial k_x},\frac{\partial a^{(\mathrm{A})}_x}{\partial k_y},\frac{\partial a^{(\mathrm{A})}_x}{\partial k_z})$ with
\begin{align}
	\partial_{k_x} a^{(\mathrm{A})}_x(\bm k) & = 2\cos k_x\sin k_z-2\sin k_y\sin k_x, \\
	\partial_{k_y} a^{(\mathrm{A})}_x(\bm k) & = 2f(\bm k)\cos k_y - 2\sin^2 k_y, \\
	\partial_{k_z} a^{(\mathrm{A})}_x(\bm k) & = 2\cos k_x\sin k_z - 2\sin k_y\sin k_z,
\end{align}
when $\nabla_{\bm k}a^{(\mathrm{A})}_x\neq 0$.
Likewise, the normal vector $\bm n^{(z)}$ of the  closed surface $a^{(\mathrm{A})}_z(\bm k)=0$ is parallel to $\nabla_{\bm k}a^{(\mathrm{A})}_z=(\frac{\partial a^{(\mathrm{A})}_z}{\partial k_x},\frac{\partial a^{(\mathrm{A})}_z}{\partial k_y},\frac{\partial a^{(\mathrm{A})}_z}{\partial k_z})$ with
\begin{align}
	\partial_{k_x} a^{(\mathrm{A})}_z(\bm k) & = 2\sin k_x \cos k_x + 2f(\bm k)\sin k_x, \\
	\partial_{k_y} a^{(\mathrm{A})}_z(\bm k) & = 2\sin k_y \cos k_y + 2f(\bm k)\sin k_y, \\
	\partial_{k_z} a^{(\mathrm{A})}_z(\bm k) & = -2\sin k_z \cos k_z + 2f(\bm k)\sin k_z, 
\end{align}
when $\nabla_{\bm k}a^{(\mathrm{A})}_z\neq 0$. Then, the tangential vector $\bm t$ of the nodal line is determined as $\bm t \parallel (\bm n^{(x)} \times \bm n^{(z)})$.

Now we focus on the touching point of the nodal chain ($m=3$), located at $\bm k=0$.
At the touching point ($\bm k=0$) in the nodal chain ($m=3$), we obtain $\nabla_{\bm k}a^{(\mathrm{A})}_x(0)=0$ and  $\nabla_{\bm k}a^{(\mathrm{A})}_z(0)=0$, which means that the tangential vector is not detemined. This is consistent with the shapes of the nodal lines, which cross perpendicularly at the touching point. Nonetheless, this result of vanishing values of $\nabla_{\bm k} a^{(\mathrm{A})}_x=(0,0,0)$ and $\nabla_{\bm k} a^{(\mathrm{A})}_z=(0,0,0)$ at the touching point does not come from physical reasons such as symmetry, but it is by accident. Since the Hamiltonian defined by Eqs. (\ref{eq:a_Ax}) and (\ref{eq:a_Az})  has only the translation and \pt\ symmetries but no other crystallographic symmetry, this result of $\nabla_{\bm k} a^{(\mathrm{A})}_x=\nabla_{\bm k} a^{(\mathrm{A})}_z=0$ cannot come from crystallographic symmetries.
Hamiltonians for real materials are complicated, and there is no reason for these two vectors to be simultaneously zero at the touching point. Therefore, the evolution of the nodal lines in this model may be unstable against perturbations.

Here we address a question whether such kinds of direct changes from nodal lines to nodal lines via nodal chains are possible. In a more general context, we study what kind of events are possible in general, which changes the topology of nodal lines.
In the next section, we  reveal what kind of events are allowed via the Morse theory.

\section{Evolutions of nodal lines without additional symmetries \label{sec:evolution_no_sym}}

\subsection{Example}

In this subsection, we show three possible topology changes in the evolutions of nodal lines in the momentum space in two-band spinless Hamiltonians with \pt\ symmetry. The three possible changes of topology of nodal lines are reconnection, annihilation, and creation. In this subsection, we give Hamiltonians to show such changes, and in the next subsection, we present a general classification scheme for these events, to show that the topology changes are restricted to the above three types.
The tangent vector of the nodal line, $\nabla_{\bm k}a_x\times\nabla_{\bm k}a_z$, is ill defined at the $\bm k$  where the topology of nodal lines changes. This is natural because the nodal lines become points at the creation and annihilation or  cross with other nodal lines at the reconnection.

\begin{figure}
	\centering
	\includegraphics[width=\linewidth]{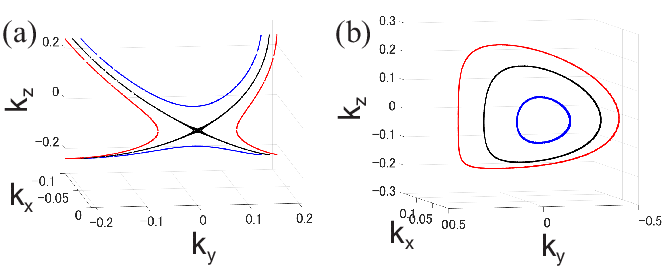}
	\caption{Possible evolutions of nodal lines in the momentum space. (a) Reconnection of nodal lines in the Hamiltonian $\mathcal{H}^{(\mathrm{B})}  (\bm k)$. The red, black, and blue lines represent nodal lines with $m=-0.003,\ 0,\text{and }0.003$, respectively. Two nodal lines touch at $m=0$, and their reconnection happen as an intermediate state from nodal lines with $m=-0.003$ to different nodal lines with $m=0.003$. (b) Annihilation (creation) of a nodal line by increasing (decreasing) the parameter $m$ in the Hamiltonian $\mathcal{H}^{(\mathrm{C})} (\bm k)$.  The red, black, and blue lines represent nodal lines with $m=-0.1,\ -0.05, \text{and } -0.01$, respectively. The size of the nodal line gets smaller when $m$ is  increased, and there are no nodal lines in $m\geq 0$.}
	\label{fig:simple_H}
\end{figure}

\subsubsection{Reconnection of nodal lines}

When a Hamiltonian $\mathcal{H}^{(\mathrm{B})}  (\bm k)$ reads
\begin{align}
	\mathcal{H}^{(\mathrm{B})}  (\bm k) & = a^{(\mathrm{B})}_x(\bm k)\sigma_x+a^{(\mathrm{B})}_z(\bm k)\sigma_z, \label{eq:H_B}  \\
	a^{(\mathrm{B})}_x(\bm k) & = \frac{1}{2}k_x^2+k_y^2+k_z^2+\frac{1}{2}m^2+k_x+2m,  \\
	a^{(\mathrm{B})}_z(\bm k) & = -\frac{1}{2}k_x^2+\frac{1}{2}k_y^2+\frac{3}{2}k_z^2-\frac{1}{2}m^2+k_x+m, 
\end{align}
a reconnection of nodal lines happens at $m=0$ as shown in Fig. \ref{fig:simple_H}(a).
The red, black, and blue lines represent nodal lines with $m=-0.003,\ 0,\text{and }0.003$, respectively. When $m$ is changed from $-0.003$ to $0.003$, two nodal lines approach each other, and they meet at $\bm k=0$ when $m=0$. Thereby they are reconnected and become two nodal lines which are different from those at $m<0$. 

\subsubsection{Annihilation of a nodal line}

When a Hamiltonian $\mathcal{H}^{(\mathrm{C})} (\bm k)$ is
\begin{align}
	\mathcal{H}^{(\mathrm{C})} (\bm k) & = a^{(\mathrm{C})}_x(\bm k)\sigma_x+a^{(\mathrm{C})}_z(\bm k)\sigma_z, \label{eq:H_C} \\
	a^{(\mathrm{C})}_x(\bm k) & = \frac{1}{2}k_x^2+k_y^2+k_z^2+\frac{1}{2}m^2+k_x+2m, \label{eq:a_Cx} \\
	a^{(\mathrm{C})}_z(\bm k) & = -\frac{1}{2}k_x^2+\frac{1}{2}k_y^2-\frac{1}{2}k_z^2-\frac{1}{2}m^2+k_x+m, \label{eq:a_Cz}
\end{align}
a nodal line is annihilated at $m=0$ as shown in Fig. \ref{fig:simple_H}(b).
The red, black, and blue lines represent nodal lines with $m=-0.1,\ -0.05, \text{and } -0.01$, respectively. When $m$ is increased from  $-0.1$ to $-0.01$, the length of the nodal line gets shorter. The nodal line shrinks to a point at $\bm k=0$, and is annihilated at $m=0$. There is no nodal line  when $m>0$.

\subsubsection{Creation of a nodal line}

The process of creating a nodal line is a reverse process of annihilating a nodal line. Hence,  the nodal line is created at $m=0$  in Fig. \ref{fig:simple_H}(b) in decreasing the parameter $m$. Alternatively, the nodal line is created at $m=0$ by increasing $m$  after a transformation $m\to -m$ in Eqs. (\ref{eq:H_C})-(\ref{eq:a_Cz}), and we label these transformed equations  as $\mathcal{H}^{(\mathrm{D})} (\bm k)$, $a^{(\mathrm{D})}_x(\bm k)$ and  $a^{(\mathrm{D})}_z(\bm k)$. Within this transformed Hamiltonian, the red, black, and blue lines in Fig. \ref{fig:simple_H}(b) correspond to nodal lines with $m=0.1,\ 0.05, \text{and } 0.01$, respectively.

We have shown  three examples for a topology change in the evolutions of nodal lines in the momentum space above and we will introduce a notion of indices in the Morse theory in the next subsection, which will be used to classify these events.

\subsection{Classification of topology changes of the nodal lines\label{sec:Classification_by_Morse_theory}}

\begin{figure}
	\centering
	\includegraphics[width=0.5\linewidth]{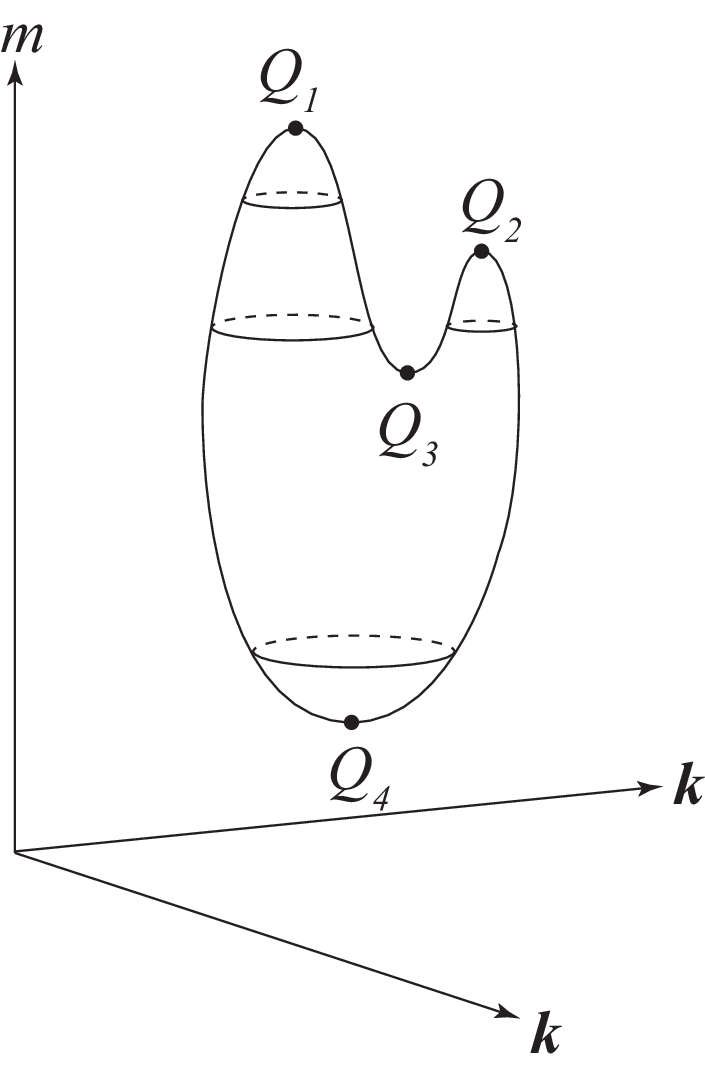}
	\caption{An example of a 2D manifold in the 4D ($\bm k, m$) space, but the 3D $\bm k$ space is described as a 2D $\bm k$ space in the figure for the sake of illustration. There are four critical points of the function $f=m$:  two local maxima at $Q_1$ and $Q_2$, a saddle point at $Q_3$, and a local minimum at $Q_4$. As the contours of this manifold  at fixed $m$ give the nodal lines, the topology of the nodal lines changes when  the value of $m$ is changed across $Q_i$ ($i=1,2,3,4$). }
	\label{fig:illust_manifold}
\end{figure}

The nodal lines evolve in the 3D $\bm k$ space with changing $m$. For the purpose of classifying the events of their topology changes, we consider a four-dimensional (4D) $(\bm k,m)$ space by adding a new axis of the parameter $m$  into the 3D $\bm k$ space, and the nodal lines in the 3D $\bm k$ space are regarded as a 2D manifold $M$ in the 4D $(\bm k,m)$ space.  In other words, conditions for nodal lines $a_x(\bm k)=0$  and $a_z(\bm k)=0$ compose the 2D manifold $M$ in the 4D $(\bm k,m)$ space because these two conditions lower the dimension by two.
Let $f$ denote a function $f:M\to\mathbb R$ giving  the value of the parameter $m$ for each point on $M$. Then, we show that the evolution of the nodal lines and their topology change are naturally described by the Morse theory \cite{Audin_2014,Nicolaescu_2011}. In the Morse theory, we define a critical point $Q$ on $M$ as a point where the gradient of $f$ is zero. In Fig. \ref{fig:illust_manifold}, we show a schematic figure of the critical point $Q$.
The  nodal lines change their topology by changing $m$ across the critical point $Q$ on the 2D manifold $M$. Each critical point $Q$ is associated with an index, which is defined as the number of negative eigenvalues in the Hessian matrix  of the function $f$ on the 2D manifold $M$. Because the Hessian matrix contains all the second-order partial derivatives of the multivariable function, its eigenvalues discriminate a local maximum, a local minimum, and a saddle point of the function $f$. Therefore  the index reveals the shape of the 2D manifold $M$ around the critical point $Q$, and characterizes the topology change of the nodal lines.
From the Morse theory we will show that the 2D manifold $M$ is allowed to have three types of shapes around a critical point $Q$ corresponding  to the three types of topology change discussed in the previous section.

In our illustrative example of a 2D manifold $M$  in Fig. \ref{fig:illust_manifold}, there are four critical points $Q_i$ ($i=1,2,3,4$) for the function $f=m$. By noting that the nodal lines are the contours at $m=\mathrm{const}$, one can see that the topology changes of the nodal lines occur at $Q_i$ by changing the value of $m$. The function $f$ has two local maxima at $Q_1$ and $Q_2$, a saddle point at $Q_3$, and a local minimum at $Q_4$, and they correspond to the three types of evolution of the nodal line, i.e., the annihilation, reconnection, and creation, respectively.

By following this scenario, we then rewrite the Hamiltonian defined in the 4D $(\bm k,m)$ space  $\mathcal{H}(\bm k) \to \mathcal{H}(\bm k,m)$, $a_{j}(\bm k) \to a_{j}(\bm k,m)$ $(j=x,z)$, and $\nabla_{\bm k}\to \nabla_{\bm k,m}\equiv\left( \frac{\partial}{\partial \bm k}, \frac{\partial}{\partial m}\right)$.
Additionally, we introduce a function $f(\bm k,m)=m$ for calculation, where $(\bm k, m)$ is a point on the 2D surface $M$.

Here, the 2D manifold $M$ is defined by the two constraints $a_x=0,a_z=0$. Therefore, from the Kamiya theorem in the Morse theory \cite{oai:soar-ir.repo.nii.ac.jp:00012243} (see Appendix \ref{app:Kamiya_theorem}), when a point $Q$ is a critical point for the function $f$ defined on the manifold $M$, $\nabla_{\bm k,m} f$ is a linear combination of $\nabla_{\bm k,m} a_x$ and $\nabla_{\bm k,m} a_z$. Namely,   the following relation is satisfied:
\begin{align}
	\nabla_{\bm k,m} f(Q)=\alpha_x\nabla_{\bm k,m} a_x(Q)+\alpha_z\nabla_{\bm k,m} a_z(Q),
\end{align}
where $\alpha_x$ and $\alpha_z$ are real parameters.
Furthermore, the index $\mathcal{N}$ of the critical point $Q$ is equal to the number of negative eigenvalues of the matrix
\begin{align}
	\mathscr{M}=\mathscr{P}(\mathscr{H}(f)|_Q-\alpha_x\mathscr{H}(a_x)|_Q-\alpha_z\mathscr{H}(a_z)|_Q)\mathscr{P}, \label{eq:Hessian}
\end{align}
where the matrix $\mathscr{P}$ represents an orthogonal projection to the 2D tangent vector space at the critical point $Q$, and $\mathscr{H}$ is a Hessian matrix.
For further discussion, we need to restrict ourselves to the cases where the critical point is nondegenerate, which means that the matrix $\mathscr{M}$ in Eq. (\ref{eq:Hessian}) has no zero eigenvalue.
Equation (\ref{eq:Hessian}) is a $2\times 2$ real symmetric matrix in terms of the basis of the tangent vector space, and so its index $\mathcal{N}$ takes the values 0, 1, and 2.
Using the Morse lemma (see Appendix \ref{app:Morse_theory}), we obtain the form of the function $f$ around the critical point $Q$ through the value of the index, and  we will see that $\mathcal{N}=2,1,0$ corresponds to annihilation, reconnection, and creation of nodal lines, respectively.
To illustrate this feature we consider the three Hamiltonians $\mathcal{H}^{(\mathrm{B})}$, $\mathcal{H}^{(\mathrm{C})}$ and $\mathcal{H}^{(\mathrm{D})}$ as examples to show how this theory works.

\subsubsection{Reconnection of nodal lines\label{sec:3B1}}

We consider the Hamiltonian $\mathcal{H}^{(\mathrm{B})}$ in Eq. (\ref{eq:H_B}) and examine the nature of the 2D surface $M$ defined by $a^{(\mathrm{B})}_j(\bm k,m)=0$ $(j=x,z)$ around its critical point $(\bm k, m)=(0,0)$  where the nodal lines are reconnected as shown  in Fig. \ref{fig:simple_H}(a). 
The gradients of $a^{(\mathrm{B})}_j(\bm k,m)$ and $f$ at $(\bm k,m)=(0,0)$ are
\begin{align}
	&\nabla_{\bm k,m}a^{(\mathrm{B})}_x(0,0)  =
	\begin{pmatrix}
		1\\
		0\\
		0\\
		2
	\end{pmatrix}, \ 
	\nabla_{\bm k,m}a^{(\mathrm{B})}_z(0,0)  =
	\begin{pmatrix}
		1\\
		0\\
		0\\
		1
	\end{pmatrix}, \label{eq:normal_vec_H_B} \\
	&\nabla_{\bm k,m}f{(0,0)}  =
	\begin{pmatrix}
		0\\
		0\\
		0\\
		1
	\end{pmatrix},
\end{align}
and we get the following relationship:
\begin{align}
	\nabla_{\bm k,m}f(0, 0)=\nabla_{\bm k,m}a^{(\mathrm{B})}_x(0,0)-\nabla_{\bm k,m}a^{(\mathrm{B})}_z(0,0).
\end{align}
This equation implies that $Q(0,0)$ is a critical point by the Kamiya theorem explained in Appendix \ref{app:Kamiya_theorem}, as expected.
Basis vectors of the space spanned by the vectors in Eq. (\ref{eq:normal_vec_H_B}) are given by
\begin{align}
	\bm b^{(\mathrm{B})}_1  =
	\begin{pmatrix}
		1\\
		0\\
		0\\
		0
	\end{pmatrix}, \ 
	\bm b^{(\mathrm{B})}_2  =
	\begin{pmatrix}
		0\\
		0\\
		0\\
		1
	\end{pmatrix},
\end{align}
and an orthogonal projection to the tangent vector space at the critical point $Q(0,0)$ is written as
\begin{align}
	\mathscr{P}^{(\mathrm{B})}   = I_4-\bm b^{(\mathrm{B})}_1\bm b^{(\mathrm{B})T}_1-\bm b^{(\mathrm{B})}_2\bm b^{(\mathrm{B})T}_2 
	 = \begin{pmatrix}
		0 &  &  & \\
		  & 1&  & \\
		  &  & 1& \\
		  &  &  & 0
	\end{pmatrix}.
\end{align}
Moreover,  Hessian matrices of $a^{(\mathrm{B})}_j(\bm k, m)$ and $f(\bm k,m)$ at the critical point are obtained as
\begin{align}
	&\mathscr{H}(a^{(\mathrm{B})}_x)|_Q=
	\begin{pmatrix}
		1 &  &  &  \\
		  & 2&  &  \\
		  &  & 2& \\
		  &  &  & 1
	\end{pmatrix}, \\
	&\mathscr{H}(a^{(\mathrm{B})}_z)|_Q=
	\begin{pmatrix}
	  -1&  &  &  \\
		& 1&  &  \\
		&  & 3&  \\
		&  &  &-1
	\end{pmatrix}, \\
	&\mathscr{H}(f)|_Q=0.
\end{align}
Therefore, the matrix $\mathscr{M}^{(\mathrm{B})}$ is written as 
\begin{align}
	\mathscr{M}^{(\mathrm{B})}& = \mathscr{P}^{(\mathrm{B})} (\mathscr{H}(f)|_Q-\mathscr{H}(a^{(\mathrm{B})}_x)|_Q+\mathscr{H}(a^{(\mathrm{B})}_z)|_Q)\mathscr{P}_\mathrm{B} \notag \\
	& = 
	\begin{pmatrix}
	   0&  &  &  \\
		&-1&  &  \\
		&  & 1&  \\
		&  &  &0
	\end{pmatrix},
\end{align}
and it indicates that the critical point $Q(0,0)$ is nondegenerate  by the Kamiya theorem  because the rank of $\mathscr{M}^{(\mathrm{B})}$ is 2, which is equal to the dimension of $M$. Furthermore, the index $\mathcal{N}$ of $f$, defined as the number of the negative eigenvalues  at the critical point $(0,0)$ is 1.

\begin{figure}
	\centering
	\includegraphics[width=\linewidth]{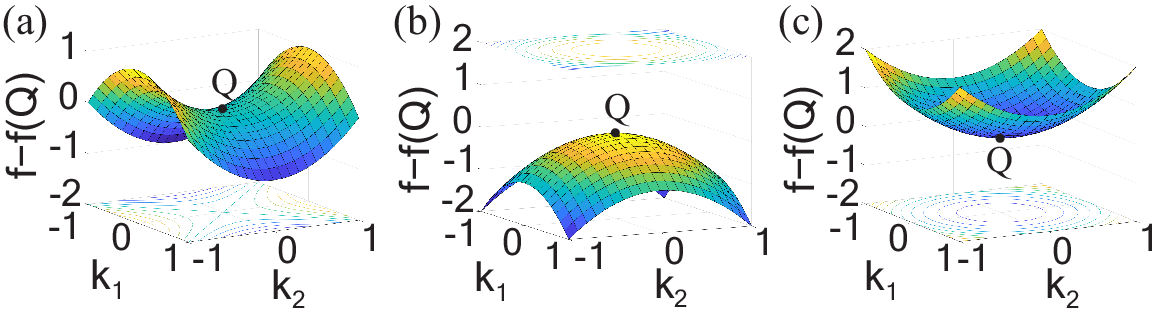}
	\caption{Relationship between an index $\mathcal{N}$ and a 2D surface $M$ around a critical point $Q$. (a) The 3D graph of $f-f(Q)=-k_1^2+k_2^2$ has a saddle point when $\mathcal{N}=1$. The 2D surface $M$ around the critical point $Q$ locally has the same shape with this graph. Reconnection of nodal lines,  which are contour lines in this figure, occurs at the critical point $Q$ when we change contour lines from bottom to top by increasing $f(=m)$. (b) The 3D graph of $f-f(Q)=-k_1^2-k_2^2$ has a local maximum when $\mathcal{N}=2$. The 2D surface $M$ around the critical point $Q$ locally has the same shape with this graph. Annihilation of nodal lines,  which are contour lines in this figure, occurs at the critical point $Q$ when we change contour lines from bottom to top by increasing $f(=m)$.  (c) The 3D graph of $f-f(Q)=k_1^2+k_2^2$ has a local minimum when $\mathcal{N}=0$. The 2D surface $M$ around the critical point $Q$ locally has the same shape with this graph. Creation of nodal lines,  which are contour lines in this figure, occurs at the critical point $Q$ when we change contour lines from bottom to top by increasing $f(=m)$.}
	\label{fig:Morse_contour}
\end{figure}

Next we discuss  the meaning of the index from the Morse lemma explained in Appendix \ref{app:Morse_theory}. Since $\mathcal{N}=1$ we can set a local coordinate ($k_1,k_2$) along $M$ around the critical point, which satisfies 
\begin{align}
	& k_1(Q)=k_2(Q)=0, \\
	& f=f(Q)-k_1^2+k_2^2.
\end{align}
This function $f$ has a saddle point, as shown in Fig. \ref{fig:Morse_contour}(a), and reproduces the shape of of the 2D manifold $M$ around the critical point $Q$. The contour lines of this function $f$ at $f=m$ are nodal lines. Thus, in Fig. \ref{fig:Morse_contour}(a) when we increase the value of $m$, the contour lines projected to the $(k_1,k_2)$ plane, i.e., the nodal lines, are reconnected across the critical point $Q$.  In that sense, a critical point $Q$ with $\mathcal{N}=1$ corresponds to the reconnection of nodal lines.

\begin{figure}
	\centering
	\includegraphics[width=0.8\linewidth]{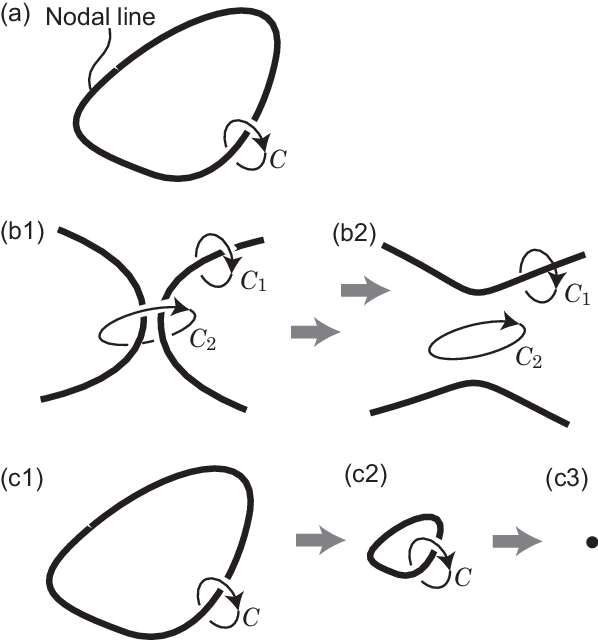}
	\caption{Schematic figures for the quantized Berry phase around the nodal lines. (a) The Berry phase around the single nodal line is quantized to be $\pi$. (b) Across the reconnection, the Berry phases along $C_1$ and $C_2$ are preserved. (c) In the annihilation of the nodal line via the change of the parameter $m$, the Berry phase along the fixed loop $C$ becomes undefined somewhere before the annihilation because the loop $C$ crosses the nodal line.}
	\label{fig:nodalline-pi}
\end{figure}

The nodal lines in this paper are characterized by the quantized $\pi$ Berry phase, and this quantization is topologically protected in systems with \pt -symmetry without SOC. This $\pi$ Berry phase is along a loop $C$ encircling the nodal line (see Fig. \ref{fig:nodalline-pi}(a)). This quantized $\pi$ Berry phase is very different from other topological invariants such as the Chern number or the $Z_2$ topological invariant, in that the Berry phase is associated with a specific loop $C$ (and therefore it depends on the choice of the loop $C$), while the Chern number and the $Z_2$ topological invariant are associated with the entire occupied bands. Thus the $\pi$ Berry phase along the loop $C$ does not tell us about any information on the phases of the topological semimetal. Therefore the topology changes of the nodal lines such as reconnection, creation, and annihilation are not related with a change of any bulk topological invariant, and they do not correspond to topological phase transition.

Meanwhile, one can argue how the Berry phase is affected by the change of topology of nodal lines. 
As we explained, the Berry phase along the loop $C$ encircling the nodal line (Fig. \ref{fig:nodalline-pi}(a)) is equal to $\pi$ (mod 2$\pi$). Along the loop $C$, the band gap is always open. Therefore, if the gap remains open on the loop $C$ under the change of the system parameter $m$, the Berry phase remains constant. For example, in Fig. \ref{fig:nodalline-pi}(b1) the Berry phase along the loop $C_1$ and $C_2$ are $\pi$ and 2$\pi$, respectively. Then if the nodal lines in Fig. \ref{fig:nodalline-pi}(b1) are reconnected, the nodal lines will look like Fig. \ref{fig:nodalline-pi}(b2), where the Berry phase along the loop $C_1$ and $C_2$ are $\pi$ and 0, respectively. Thus, considering that the Berry phase is defined in terms of modulo $2\pi$, the Berry phase is unaffected by the reconnection. It is also seen in Fig. \ref{fig:simple_H}. The Berry phase along a closed loop on the $k_z$=0 plane encircling the two nodal lines (red lines) at $m= -0.003$ in Fig. \ref{fig:simple_H}(a). Via the change of $m$ through $m=0$ (the change from red to blue lines in Fig. \ref{fig:simple_H}(a)), the Berry phase changes from 2$\pi$ to 0 (modulo 2$\pi$), which means that the Berry phase remains constant across the reconnection. It is natural because on the closed path, the system remains gapped and therefore the Berry phase cannot have a jump across the reconnection.

On the other hand, suppose the nodal line in Fig. \ref{fig:nodalline-pi}(c1) is annihilated as shown in Figs. \ref{fig:nodalline-pi}(c2) and \ref{fig:nodalline-pi}(c3). In this case, for the fixed loop $C$, it will eventually cross the nodal line {\it before} the nodal line shrinks to a point, and the Berry phase jumps from $\pi$ to 0. The value of $m$ where this jump occurs depends on the position of the loop $C$, and it does not correspond to the value of $m$ where the nodal line is annihilated. Thus, the Berry phase for the specific loop $C$ does not give any information on the topological phase.

\subsubsection{Annihilation of a nodal line}

We examine the nature of the 2D surface $M$ defined by $a^{(\mathrm{C})}_j(\bm k,m)=0$ $(j=x,z)$ around its critical point $(\bm k, m)=(0,0)$  where the nodal line is annihilated as shown  in Fig. \ref{fig:simple_H}(b). 
By the classification similar to Sec. \ref{sec:3B1}, the point  $Q:(\bm k,m)=(0,0)$ is a critical point of the function $f$ with the index $\mathcal{N}=2$.

From the Morse lemma, we can set a local coordinate ($k_1,k_2$) around the critical point, which satisfies 
\begin{align}
	& k_1(Q)=k_2(Q)=0, \\
	& f=f(Q)-k_1^2-k_2^2.
\end{align}
This function $f$ means a local maximum of the 2D manifold $M$ around the critical point $Q$, as shown in Fig. \ref{fig:Morse_contour}(b). When we increase the value of $m$, the contour line projected to the $(k_1,k_2)$ plane  vanishes.  Hence, a critical point $Q$ with $\mathcal{N}=2$ leads to the annihilation of nodal lines.

\subsubsection{Creation of a nodal line}

We examine the nature of the 2D surface $M$ defined by $a^{(\mathrm{D})}_j(\bm k,m)=0$  $(j=x,z)$ around the critical point $(\bm k, m)=(0,0)$, having the index $\mathcal{N}=0$, where the nodal line is created as shown  in Fig. \ref{fig:simple_H}(c).

From the Morse lemma, we can set a local coordinate ($k_1,k_2$) around the critical point, which satisfies
\begin{align}
	& k_1(Q)=k_2(Q)=0, \\
	& f=f(Q)+k_1^2+k_2^2.
\end{align}
This function $f$ has a local minimum around the critical point $Q$, as shown in Fig. \ref{fig:Morse_contour}(c). The change of the contour line projected to the $(k_1,k_2)$ plane obtained by increasing the value of $m$ reveals the creation of the nodal line. Thus, a critical point $Q$ with $\mathcal{N}=0$ indicates the creation of nodal lines.

\subsection{Results of a speciality removed model for the nodal link}
As we discussed in Sec. \ref{sec:previous-nodal-link}, in the model described by Eqs. (\ref{eq:H_A})-(\ref{eq:a_Az}), both $\nabla_{\bm k}a^{(A)}_x$ and $\nabla_{\bm k}a^{(A)}_z$ vanish at the touching point of the nodal chain. 
This does not come from physical reasons such as symmetry, and is considered as an artifact of the special choice of the model. We can remove this  artifact by adding some terms to the Hamiltonian. For example, we redefine the Hamiltonian by $a^{(A)}_x\to a'^{(A)}_x=a^{(A)}_x+\alpha\sin k_x$ and $a^{(A)}_z\to a'^{(A)}_z=a^{(A)}_z+\alpha\sin k_z$, where $\alpha$ is a real parameter. Then, as shown in Appendix \ref{app:further_explanation_nodal_link}, instead of the direct transition from the nodal lines to a nodal chain, a reconnection of nodal lines occur three times to get a nodal chain. 

In fact, this is expected from the Morse theory in this section. If we assume no crystallographic symmetry except for translation and \pt\ symmetries, the discussion in this  section section tells us that the topology change of nodal lines are restricted to three types, creation, reconnection, and annihilation. Meanwhile, the model (\ref{eq:H_A})-(\ref{eq:a_Az}) in Sec. \ref{sec:NL_with_pi_berry_pahse} exhibits a direct transition from nodal lines to a nodal link, and this event is not among the three types described above. This means that the model (\ref{eq:H_A})-(\ref{eq:a_Az}) is not general but specially designed, and this transition from nodal lines to a nodal link is unstable against perturbations. Namely, from nodal lines to a nodal link, one cannot have direct transition, but it can be realized through multiple reconnections of nodal lines, if no additional crystallographic symmetries are assumed.

\subsection{Short summary of this section}

\begin{figure}
	\centering
	\includegraphics[width=\linewidth]{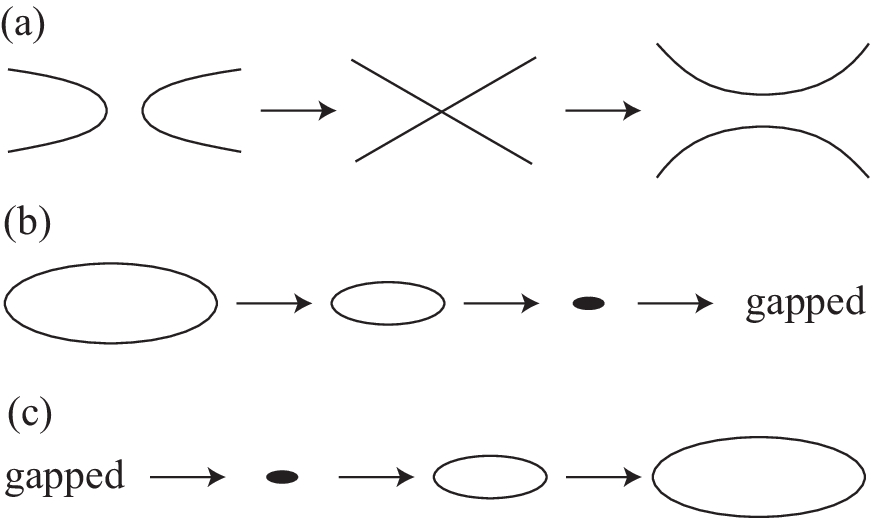}
	\caption{Evolution of nodal lines with \pt\ symmetry in the momentum space. (a) The reconnection of nodal lines happen in the middle panel. (b) The nodal line is annihilated from the left panel to the right panel. (c) The nodal line is created from the left panel to the right panel.}
	\label{fig:illust_simple}
\end{figure}

In this section, we reveal that there are three possible events of topology change of nodal lines from the above calculations. These events are characterized by the index $\mathcal{N}$ from  the Morse theory, and $\mathcal{N}=2$, $\mathcal{N}=1$, and $\mathcal{N}=0$ corresponds to  the annihilation, the reconnection, and the creation of nodal lines, as schematically shown in Figs. \ref{fig:illust_simple}(a), \ref{fig:illust_simple}(b), and \ref{fig:illust_simple}(c), respectively, where $f(=m)$ is a parameter driving the evolution of nodal lines.  The topology change of the nodal line in the vicinity of the critical point is illustrated in terms of the 2D local coordinates around the critical point in Fig. \ref{fig:Morse_contour}.

\section{Evolutions of nodal lines in systems with additional symmetries \label{sec:evolution_with_sym}}

This section shows topology changes of nodal lines with additional crystallographic symmetries in \pt -symmetric systems. For this purpose, we characterize the topology change in terms of the index $\mathcal{N}$ in the Morse theory, by  using $\bm k\cdot\bm p$ models with rotational or mirror symmetry.

We start with the $2\times 2$ Hamiltonian $\mathcal{H}(\bm k)=a_x(\bm k)\sigma_x+a_z(\bm k)\sigma_z$ with \pt\ symmetry. In the presence of other crystallographic symmetries, we focus on a topology change of nodal lines at the $\bm k$ point invariant under this crystallographic symmetry, and let $\mathcal{G}$ denote the little group at this $\bm k$ point. Then, the $2\times 2$  effective Hamiltonian around that point satisfies
\begin{align}
	D(g)\mathcal{H}(\bm k)D^{-1}(g)=\mathcal{H}(g\bm k) \quad  \forall g \in{\mathcal{G}}, \label{eq:theory_of_invariants}
\end{align}
where $D(g)$ is a representation matrix of the symmetry operation $g\in{\mathcal{G}}$. In the following, we consider the cases with $C_n$ symmetry and with mirror symmetry.

\subsection{Cases with $C_n$ symmetries ($n=2,3,4,6$)\label{sec:C_n}}

Among various rotational symmetries, only the $n$-fold rotational $(C_n)$ symmetry with $n=2,3,4,6$ is allowed in crystals. We show how a nodal line evolves under the rotational symmetries.

\subsubsection{$C_2$ symmetry}

\begin{figure}
	\centering
	\includegraphics[width=\linewidth]{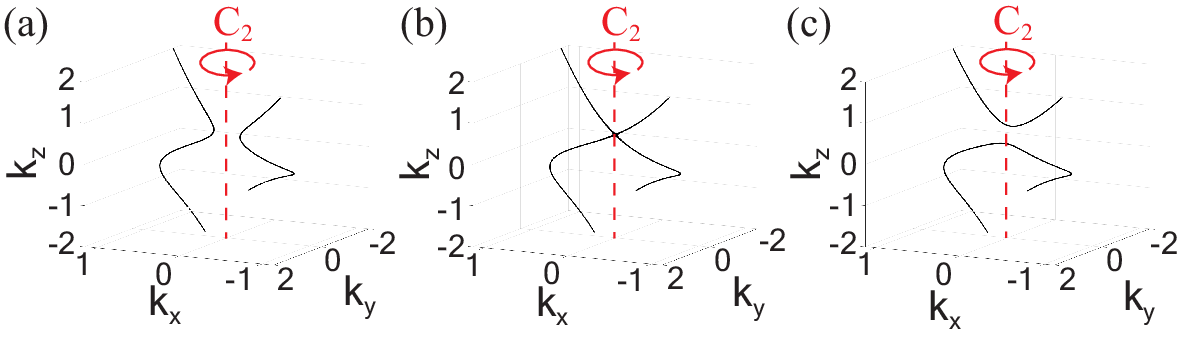}
	\caption{Evolutions of nodal lines with the $C_2$  symmetry in the momentum space. Nodal lines are given by the Hamiltonian  $\mathcal{H}^{(C_2)}  (\bm k)$ with $c_1=1,c_2=\frac{1}{2},c_3=\frac{1}{4},c_4=\frac{1}{2},c_5=1,c_6=1,c_7=-1,c_8=1,c_9=1$. (a), (b) and (c) represent nodal lines with $m=-0.3,-0.25,-0.2$, respectively. The nodal lines are reconnected at $m=-0.25$ in (b).}
	\label{fig:Two-fold-rotation}
\end{figure}

We consider nodal lines with \pt\ symmetry and two-fold rotational symmetry  with respect to the $k_z$ axis.
In spinless systems, the eigenvalue of $C_2$ is 1 for the irreducible representation (irrep) A and $-1$ for the irrep B, and we use conventional names for irreps known as Mulliken symbols \cite{PhysRev.43.279,BB01897854}.  We discuss topology changes of nodal lines at a point $\bm k=\bm k_0$ on the  $C_2$ axis ($k_z$ axis). Let us consider the case where one of the two bands follows the irrep A and the other follows the irrep B. Later in Sec. \ref{sec:Cn_with_same_irrep}, we consider the case with the two bands following the same irreps. Then, a representation matrix of the $C_2$ rotation  is obtained as
\begin{align}
	D(C_2)=
	\begin{pmatrix}
		1 & 0  \\
		0 & -1 
	\end{pmatrix}.
\end{align}
We shift the origin in $\bm k$ space by $\bm k_0$, so that the focused point becomes $\bm k=0$.
By using Eq. (\ref{eq:theory_of_invariants}), the $\bm k\cdot\bm p$ Hamiltonian around $\bm k=0$ is written up to the second order in $\bm k$ as
\begin{align}
	\mathcal{H}^{(C_2)}(\bm k)= & a^{(C_2)}_x\sigma_x+a^{(C_2)}_z\sigma_z, \label{eq:H_C2} \\
	a^{(C_2)}_x(\bm k)= & c_1k_xk_z+c_2k_yk_z+c_3k_x+c_4k_y, \label{eq:a_c2x} \\
	a^{(C_2)}_z(\bm k)= & c_5k_x^2+c_6k_y^2+c_7k_z^2 +c_8k_xk_y+c_9k_z+m, \label{eq:a_c2z}
\end{align}
where $c_i \quad (i=1,2,\cdots ,9)$ and $m$ are real parameters. We set $c_1=1,c_2=\frac{1}{2},c_3=\frac{1}{4},c_4=\frac{1}{2},c_5=1,c_6=1,c_7=-1,c_8=1,c_9=1$ as an example. The parameter $m$, which plays a role of driving the evolution of nodal lines, is introduced as a constant term in Eq. (\ref{eq:a_c2z}) for simplicity. As shown in Fig. \ref{fig:Two-fold-rotation}, a reconnection of nodal lines happens. Figures \ref{fig:Two-fold-rotation}(a), \ref{fig:Two-fold-rotation}(b) and \ref{fig:Two-fold-rotation}(c) represent nodal lines with $m=-0.3,-0.25,-0.2$, respectively. When $m$ is changed from $m=-0.3$ to $-0.2$, the nodal lines are reconnected at $\bm k=(0,0,\frac{1}{2})$ with $m=-\frac{1}{4}$. By the calculation simillar to Sec. \ref{sec:Classification_by_Morse_theory}, we find that the point $(\bm k, m)=(0,0,\frac{1}{2},-\frac{1}{4})$ is a critical point with index $\mathcal{N}=1$ by the Kamiya theorem. Then the reconnection of the nodal lines must occur by the Morse lemma, in agreement with Fig. \ref{fig:Two-fold-rotation}.

\begin{figure}
	\centering
	\includegraphics[width=\linewidth]{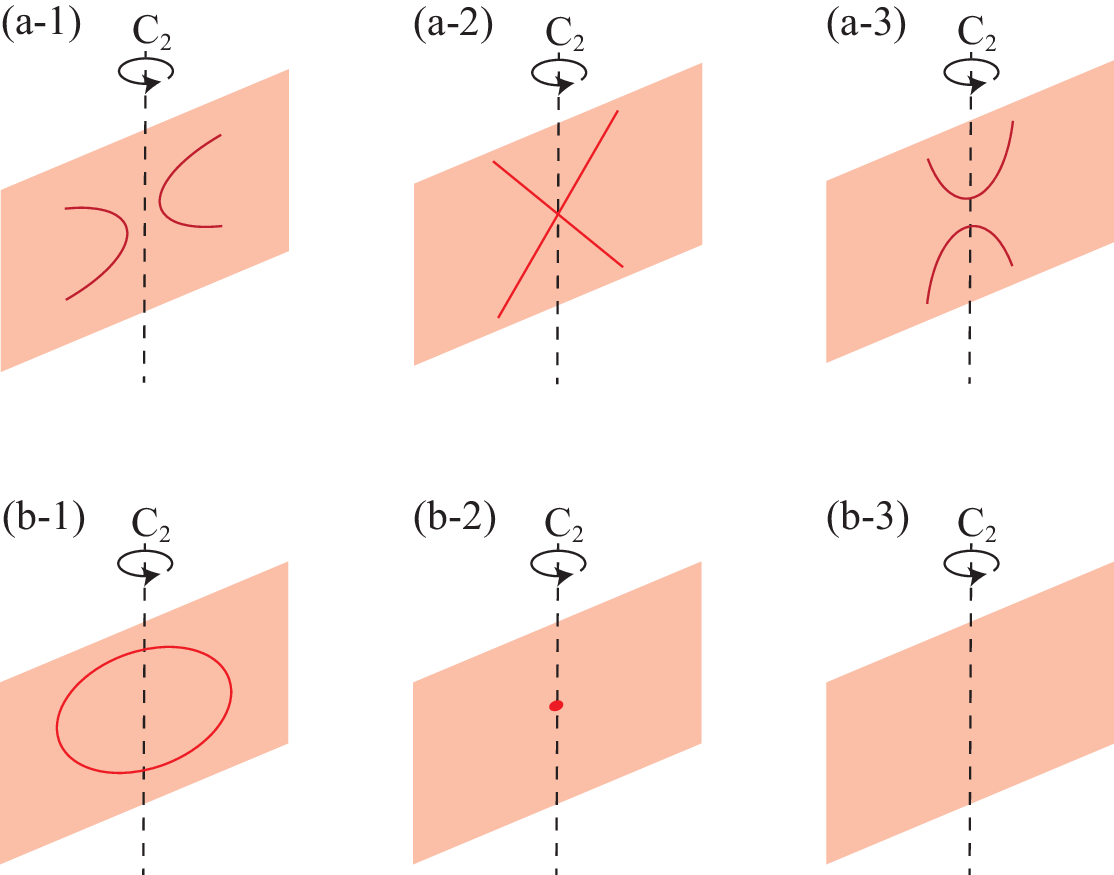}
	\caption{Schematic pictures for the topology changes of nodal lines with the two-fold rotational symmetry in the momentum space. The red colored nodal lines reside inside the $C_2$-symmetric red plane near the $C_2$ axis. The reconnection of nodal lines occurs in (a-1)-(a-3), and the annihilation or creation occur in (b-1)-(b-3). We note that away from the $C_2$ axis, the nodal lines can be away from the red plane. }
	\label{fig:illust_two-roto}
\end{figure}

Next, we generalize the result on the model Hamiltonian (\ref{eq:H_C2}), and consider  $a_x(\bm k)$ and $a_z(\bm k)$ as general analytic functions of $\bm k$.
Because of Eq. (\ref{eq:theory_of_invariants}), they follow $a_x(k_x,k_y,k_z)=-a_x(-k_x,-k_y,k_z)$ and $a_z(k_x,k_y,k_z)=a_z(-k_x,-k_y,k_z)$. Then, the gradients of the functions $f$, $a_x$, and $a_z$ for the Morse theory at a point on the $C_2$ axis ($k_z$ axis) are obtained as
\begin{align}
	\nabla_{\bm k,m} f(0,0,k_z,m)=(0,0,0,1)^T,  \label{eq:nabla_f_c2}\\
	\nabla_{\bm k,m} a_x(0,0,k_z,m)=(\cdot,\cdot,0,0)^T, \label{eq:nabla_a_c2x} \\
	\nabla_{\bm k,m} a_z(0,0,k_z,m)=(0,0,\cdot,\cdot)^T  \label{eq:nabla_a_c2z},
\end{align}
where $\cdot$ represents a term left undetermined only from the symmetry, and such a term is in general nonzero. 
If there is a point $(0,0,k_z,m)$ where Eqs. (\ref{eq:nabla_f_c2})-(\ref{eq:nabla_a_c2z}) are linearly dependent,  the point is a critical point by the Kamiya theorem. It occurs when $\frac{\partial a_z}{\partial k_z}=0$, leading $\nabla_{\bm k,m}f \propto \nabla_{\bm k,m}a_z$. 
Therefore, the critical point $(\bm k,m)=(0,0,k_z,m)$ is determined by two conditions, $a_z=0$ and $\frac{\partial a_z}{\partial k_z}=0$, because on the $C_2$ axis, $a_x(0,0,k_z,m)$ vanishes because of symmetry. Since the number of equations is equal to the number of variables $(k_z,m)$, they can have solutions which are isolated points in the $(\bm k,m)$ space in general.  Then, the eigenvalues of the $2\times 2$ matrix $\mathscr{M}$ are nonzero in general, and the critical point is classified in terms of the index $\mathcal{N}$ into three cases, $\mathcal{N}=0,1,2$, corresponding to creation, reconnection, and annihilation of nodal lines, respectively.

In order to see how the nodal lines evolve under $C_2$ symmetry, we note that the vector $\nabla_{\bm k}a_x|_P=(\cdot,\cdot,0)$ at the critical point $P$ on the $C_2$ axis defines a normal vector of the surface $a_x=0$. Thus, in the vicinity of the critical point $P$, the nodal lines evolve along the plane normal to this vector $\nabla_{\bm k}a_x|_P=(\cdot,\cdot,0)$.
This plane is  $C_2$-symmetric and it contains the $C_2$  axis  as shown in Fig. \ref{fig:illust_two-roto}. The red lines represent nodal lines, which lie along the red plane containing $C_2$ axis, and the nodal lines follow $C_2$ symmetry. 
For example, when $\mathcal{N}=1$ and the reconnection  happens, the nodal lines evolve from Figs. \ref{fig:illust_two-roto}(a-1) to \ref{fig:illust_two-roto}(a-3) through \ref{fig:illust_two-roto}(a-2), and vice versa.  This result from symmetry and the Morse theory matches with that  of the numerical calculation in Fig. \ref{fig:Two-fold-rotation} in the vicinity of the critical point.
Next when $\mathcal{N}=2$ and the annihilation happens, the nodal lines evolve from Figs. \ref{fig:illust_two-roto}(b-1) to \ref{fig:illust_two-roto}(b-3) through \ref{fig:illust_two-roto}(b-2). The nodal ring shrinks to a point on $C_2$ axis as shown in Fig. \ref{fig:illust_two-roto}(b-2). Then, the case with $\mathcal{N}=0$ corresponds to the creation of nodal lines, and is a reverse process, from Figs. \ref{fig:illust_two-roto}(b-3) to \ref{fig:illust_two-roto}(b-1).

\subsubsection{$C_3$ symmetry}
We consider nodal lines with \pt\ symmetry and three-fold rotational symmetry  with respect to the $k_z$ axis.
In spinless systems, the eigenvalue of $C_3$ is 1 for the irrep A, $\mathrm{e}^{2\pi i/3}$ for the irrep $^2\mathrm{E}$, and $\mathrm{e}^{-2\pi i/3}$ for the irrep $^1\mathrm{E}$. Under the \pt\ symmetry, the irreps $^1\mathrm{E}$ and $^2\mathrm{E}$ are degenerate because they are complex representations, and a nodal line formed by $^1\mathrm{E}$ and $^2\mathrm{E}$ irreps always lies along the $C_3$ axis. Therefore, it is impossible to see the topology change reflecting $C_3$ symmetry, and we can exclude this case from our discussion.

\subsubsection{$C_4$ symmetry\label{sec:Result_C4}}

\begin{figure}
	\centering
	\includegraphics[width=\linewidth]{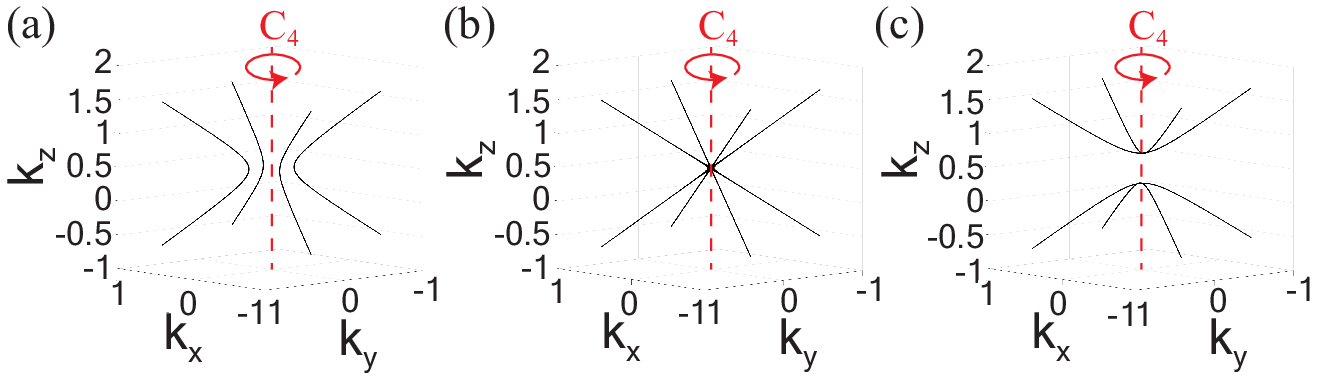}
	\caption{Evolutions of nodal lines with the four-fold rotational symmetry in the momentum space. Nodal lines are given by the Hamiltonian  $\mathcal{H}^{(C_4)}  (\bm k)$ with $c_1=1,c_2=2,c_3=1,c_4=-1,c_5=1$. (a), (b) and (c) represent nodal lines with $m=-0.3,-0.25,-0.2$, respectively. The nodal lines are reconnected at $m=-0.25$ in (b). }
	\label{fig:Four-fold-rotation}
\end{figure}

We consider nodal lines with \pt\ symmetry and four-fold rotational symmetry  with respect to the $k_z$ axis.
We discuss topology changes of nodal lines at a point $\bm k=\bm k_0$ on the $C_4$ axis ($k_z$ axis).
In spinless systems, the eigenvalue of $C_4$ is 1 for the irrep A, $-1$ for the irrep B, $i$ for the irrep $^2\mathrm{E}$, and $-i$ for the irrep $^1\mathrm{E}$. We can exclude the irreps $^1\mathrm{E}$ and $^2\mathrm{E}$ from our discussion because under the \pt\ symmetry the complex irreps $^1\mathrm{E}$ and $^2\mathrm{E}$ are degenerate. Therefore, let us  consider the case where one of the two bands follows the irrep A and the other follows the irrep B. The cases with the two bands following the same irrep A (or B) will be discussed in Sec. \ref{sec:Cn_with_same_irrep}. Then, the representation matrix for $C_4$  is obtained as
\begin{align}
	D(C_4)=
	\begin{pmatrix}
		1 & 0  \\
		0 & -1 
	\end{pmatrix}. \label{eq:D_C4}
\end{align}
We focus on a point $\bm k_0$ on the $C_4$ axis. For convenience we shift the origin in $\bm k$ space by $\bm k_0$, so that the focused point becomes $\bm k=0$.

Now we discuss the evolution of nodal lines under the $C_4$ symmetry with Eq. (\ref{eq:D_C4}). We begin with a simple example; by using Eq. (\ref{eq:theory_of_invariants}), the $\bm k\cdot\bm p$ Hamiltonian around $\bm k=0$ is written up to the second order in $\bm k$ as
\begin{align}
	\mathcal{H}^{(C_4)}(\bm k)= & a^{(C_4)}_x\sigma_x+a^{(C_4)}_z\sigma_z, \label{eq:H_C4} \\	
	a^{(C_4)}_x(\bm k)= & c_1(k_x^2-k_y^2)+c_2k_xk_y, \\
	a^{(C_4)}_z(\bm k)= & c_3(k_x^2+k_y^2)+c_4k_z^2+c_5k_z+m, \label{eq:a_c4z}
\end{align}
where $c_i \quad (i=1,2,\cdots ,5)$ and $m$ are real parameters. We set $c_1=1,c_2=2,c_3=1,c_4=-1,c_5=1$ as an example.
We show the result in Fig. \ref{fig:Four-fold-rotation}, where nodal lines are reconnected by increasing $m$. Figures \ref{fig:Four-fold-rotation}(a), \ref{fig:Four-fold-rotation}(b), and \ref{fig:Four-fold-rotation}(c) represent nodal lines with $m=-0.3,-0.25,-0.2$, respectively. When $m$ is changed from $m=-0.3$ to $m=-0.2$, the nodal lines are reconnected at $\bm k=(0,0,\frac{1}{2})$ with $m=-\frac{1}{4}$.
Next, we want to characterize the point $P$ $(\bm k,m)=(0,0,\frac{1}{2},-\frac{1}{4})$ by the Morse theory. Nevertheless, within our scenario in the previous sections, this point is not a  critical point in the Morse theory since $\nabla_{\bm k}a_x^{(C_4)}=0$ at this point. 
As a result, one cannot study this reconnection in terms of the Morse theory.

Here we find that by adopting the following argument of factorization this point $P$ can be regarded as a nondegenerate critical point in the Morse theory, and then the classification in terms of the index $\mathcal{N}$ can now be used. First, we note that $a^{(C_4)}_x$ can be factorized under the $C_4$ symmetry:
\begin{align}
	a^{(C_4)}_x(\bm k) & =((\sqrt{2}-1)k_x+k_y)((\sqrt{2}+1)k_x-k_y)  \notag \\
	& \equiv a^{(C_4)(\mathrm{I})}_x(\bm k)a^{(C_4)(\mathrm{I}\hspace{-1pt}\mathrm{I})}_x(\bm k) \label{eq:a_c4x_factorized},
\end{align}
where $a^{(C_4)(\mathrm{I})}_x(\bm k)=(\sqrt{2}-1)k_x+k_y$ and $a^{(C_4)(\mathrm{I}\hspace{-1pt}\mathrm{I})}_x(\bm k)=(\sqrt{2}+1)k_x-k_y$. Equation (\ref{eq:a_c4x_factorized}) means that $a^{(C_4)}_x(\bm k)$ changes its sign four times around the $k_z$ axis as is expected from the symmetry constraint $a^{(C_4)}_x(\bm k)=-a^{(C_4)}_x(C_4\bm k)$.
Therefore, the condition for nodal lines is decomposed as
\begin{enumerate}[label=(\Roman*),align=parleft,leftmargin=50pt,ref=(\Roman*),rightmargin=20pt]
	\item $a^{(C_4)(\mathrm{I})}_x(\bm k)=0$ and $a^{(C_4)}_z(\bm k)=0$ ,
	\item $a^{(C_4)(\mathrm{I}\hspace{-1pt}\mathrm{I})}_x(\bm k)=0$ and $a^{(C_4)}_z(\bm k)=0$.
\end{enumerate}
We  discuss a change of topology of nodal lines in the cases (I) and (I\hspace{-1pt}I) separately.
In both  cases (I) and (I\hspace{-1pt}I), the point $(\bm k,m)=(0,0,\frac{1}{2},-\frac{1}{4})$ now becomes a critical point with index $\mathcal{N}=1$ by the Kamiya theorem. At this critical point, the reconnections of nodal lines occur on $a^{(C_4)(\mathrm{I})}_x(\bm k)=(\sqrt{2}-1)k_x+k_y=0$ and $a^{(C_4)(\mathrm{I}\hspace{-1pt}\mathrm{I})}_x(\bm k)=(\sqrt{2}+1)k_x-k_y=0$ planes  by the Morse lemma.
The change of topology of nodal lines occurs in a $C_4$-symmetric way because cases (I) and  (I\hspace{-1pt}I) are related to each other by $C_4$ symmetry.

\begin{figure}
	\centering
	\includegraphics[width=\linewidth]{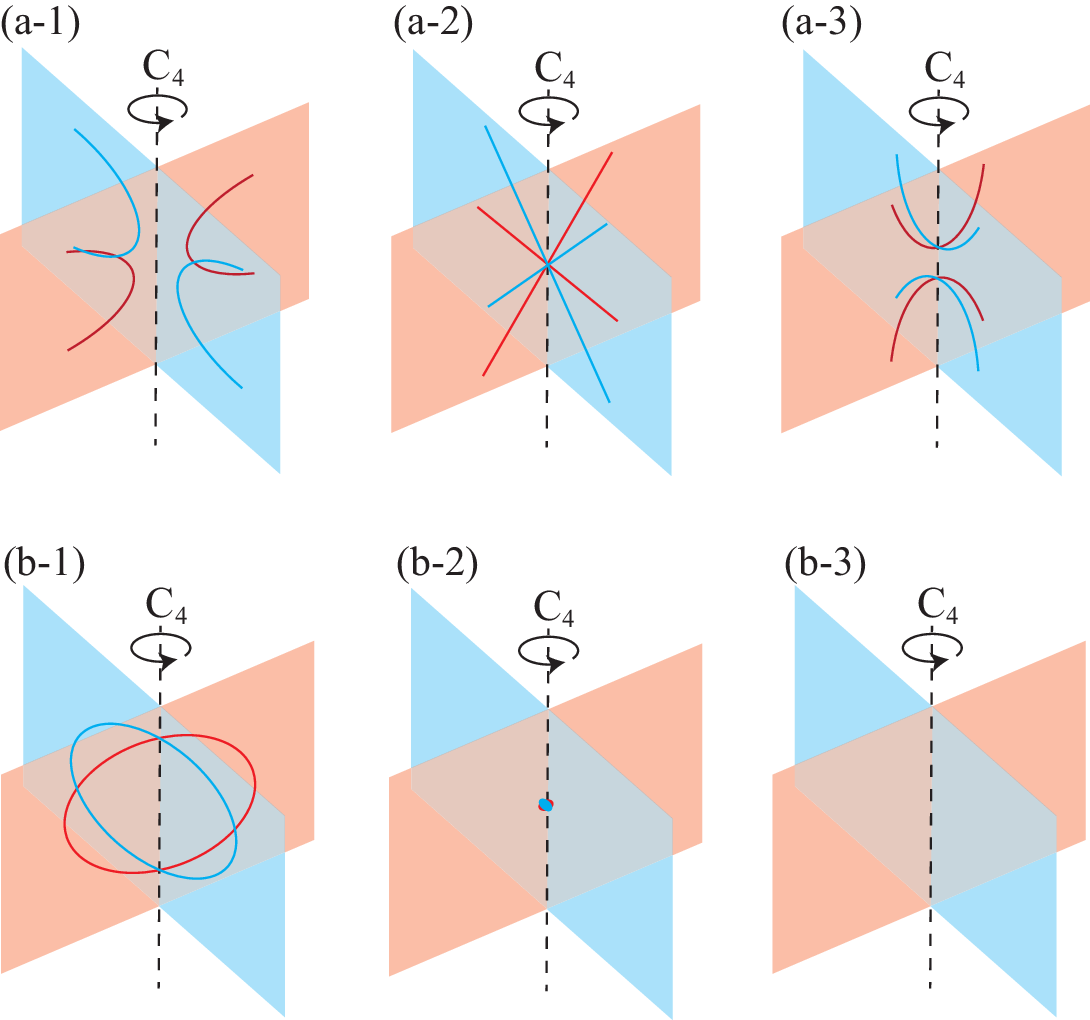}
	\caption{Schematic pictures of the topology changes of nodal lines with the four-fold rotational symmetry in the momentum space. The red (blue) colored line represents the nodal line inside the red (blue) plane near the $C_4$ axis, which holds $C_4$ symmetry in total. There are the reconnection of the nodal lines in (a-1)-(a-3), and the annihilation (creation) of the nodal lines in (b-1)-(b-3). Away from the $C_4$ axis, the nodal lines may leave the blue and red planes.}
	\label{fig:illust_four-roto}
\end{figure}

Even apart from the example in Eqs. (\ref{eq:H_C4})-(\ref{eq:a_c4z}), in general $C_4$-symmetric systems we can show that the topology change of nodal lines is fully characterized by the index $\mathcal{N}$ for critical points. As seen in the above example, the key finding is that $a_x(\bm k)$ can always be factorized under the $C_4$ symmetry. Then, by using each factor of $a_x$ (not $a_x$ itself), the points of the topology change of nodal lines become nondegenerate critical points in the Morse theory.
To see this we assume that $a_x(\bm k)$ and $a_z(\bm k)$ are analytic in $\bm k$ around the critical point $(0,0,k_z^{(0)})$, and can generally contain higher order terms in $\bm k$. Even then, we show that $a_x(\bm k)$ is factorized:
\begin{align}
	a_x(\bm k)&=i(\alpha k_{+}g(\bm k)-\bar\alpha k_{-}\bar{g}(\bm k))(\alpha k_{+}g(\bm k)+\bar\alpha k_{-}\bar{g}(\bm k)), \notag \\
	&\equiv ia^{(-)}_x(\bm k)a^{(+)}_x(\bm k), \label{eq:a_c4x_factorized_general}
\end{align}
where $a^{(\pm)}_x(\bm k)=\alpha k_{+}g(\bm k)\pm\bar\alpha k_{-}\bar{g}(\bm k)$, $k_\pm=k_x\pm ik_y$, and $\alpha$ is a complex constant. The function $g(\bm k)$ is analytic in $\bm k$ and satisfies $g(0,0,k_z^{(0)})=1$ and $g(\bm k)=g(C_4\bm k)$. The detailed proof of Eq. (\ref{eq:a_c4x_factorized_general}) is in Appendix \ref{app:C4}. Hence, nodal lines appear in the following cases:
\begin{enumerate}[label=(\Roman*),align=parleft,leftmargin=50pt,ref=(\Roman*),rightmargin=20pt]
	\item $a^{(-)}_x(\bm k)=0$ and $a_z(\bm k)=0$ ,
	\item $a^{(+)}_x(\bm k)=0$ and $a_z(\bm k)=0$.
\end{enumerate}
Because $a^{(\pm)}_x(C_4\bm k)=ia^{(\mp)}_x(\bm k)$, $a^{(\pm)}_x(C_2\bm k)=-a^{(\pm)}_x(\bm k)$, and $a^{(\pm)}_x(0,0,k_z^{(0)})=0$, each of the equations $a^{(\pm)}_x(\bm k)=0$ represents a $C_2$-symmetric surface containing the rotational axis, and the combination of the two surfaces is  $C_4$ symmetric. 

Now, we apply the Morse theory in (I) and (I\hspace{-1pt}I) separately.
We obtain the gradients of the functions used in the Morse theory at a point along the $C_4$ axis ($k_z$ axis):
\begin{align}
	\nabla_{\bm k,m} f(0,0,k_z,m)=(0,0,0,1)^T, \\
	\nabla_{\bm k,m} a^{(\pm)}_x(0,0,k_z,m)=(\cdot,\cdot,0,0)^T,  \\
	\nabla_{\bm k,m} a_z(0,0,k_z,m)=(0,0,\cdot,\cdot)^T.
\end{align}
In case (I) (and also in case (I\hspace{-1pt}I)), similar to the $C_2$-symmetric case the critical point should satisfy  $\nabla_{\bm k,m}f \propto \nabla_{\bm k,m}a_z$. 
Thus the critical point $(\bm k,m)=(0,0,k_z,m)$ is determined by two conditions $a_z=0$ and $\frac{\partial a_z}{\partial k_z}=0$, and this set of equations can have solutions for two variables $k_z$ and $m$. Because these conditions, $a_z=0$ and $\frac{\partial a_z}{\partial k_z}=0$, are common between cases (I) and (I\hspace{-1pt}I), the critical points are common.

The resulting evolutions of nodal lines across the topology change are illustrated in Fig. \ref{fig:illust_four-roto}.
The normal vectors of the two surfaces $a^{(\pm)}_x(\bm k)=0$ are given by $\nabla_{\bm k}a^{(\pm)}_x(\bm k)$; hence at the critical point, they are perpendicular to each other and also perpendicular to the $z$-axis.
Evolutions of nodal lines are confined to each plane near the critical point, and they follow the $C_4$ symmetry in total.
When $\mathcal{N}=1$ corresponding to the reconnection, the nodal lines evolve from Figs. \ref{fig:illust_four-roto}(a-1) to \ref{fig:illust_four-roto}(a-3) through \ref{fig:illust_four-roto}(a-2), and vice versa, which is in  a good agreement with the result by the numerical calculation in Fig. \ref{fig:Four-fold-rotation}.
When $\mathcal{N}=2$ and the annihilation happens, the nodal lines evolve from Figs. \ref{fig:illust_four-roto}(b-1) to \ref{fig:illust_four-roto}(b-3) through \ref{fig:illust_four-roto}(b-2). Its reverse process from Figs. \ref{fig:illust_four-roto}(b-3) to \ref{fig:illust_four-roto}(b-1) corresponds to the creation of a nodal line with $\mathcal{N}=0$.

\subsubsection{$C_6$ symmetry\label{sec:IVA4}}

\begin{figure}
	\centering
	\includegraphics[width=\linewidth]{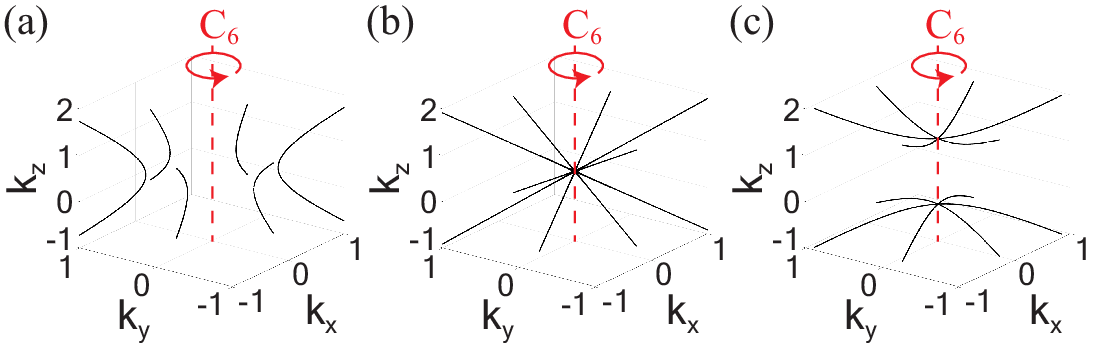}
	\caption{Evolutions of nodal lines with the four-fold rotational symmetry in the momentum space. Nodal lines are given by the Hamiltonian  $\mathcal{H}^{(C_6)}  (\bm k)$ with $c_1=1,c_2=1,c_3=0,c_4=0,c_5=1,c_6=-1,c_7=1$. (a), (b), and (c) represent nodal lines with $m=-0.75,-0.25,0.25$, respectively. The nodal lines are reconnected at $m=-0.25$ in (b). }
	\label{fig:Six-fold-rotation}
\end{figure}

We consider nodal lines with \pt\ symmetry and six-fold rotational symmetry  along the $k_z$ axis.
We discuss topology changes of nodal lines at a point $\bm k=\bm k_0$ on the $C_6$ axis ($k_z$ axis).
In spinless systems, the eigenvalue of $C_6$ is 1 for the irrep A, $-1$ for the irrep B, $\mathrm{e}^{-2\pi i/3}$ for the irrep $^2\mathrm{E}_1$, $\mathrm{e}^{\pi i/3}$ for the irrep $^2\mathrm{E}_2$, $\mathrm{e}^{2\pi i/3}$ for the irrep $^1\mathrm{E}_1$, and $\mathrm{e}^{-\pi i/3}$ for the irrep $^1\mathrm{E}_2$. Here, the pairs of irreps $(^1\mathrm{E}_{1},^2\mathrm{E}_{1})$ and $(^1\mathrm{E}_{2},^2\mathrm{E}_{2})$ form Kramers degeneracy because they are complex representations. Therefore we can restrict ourselves to the irreps A and B. We will study the case where the two bands follow the same irreps (A or B) in Sec. \ref{sec:Cn_with_same_irrep}. In this  subsection we study the remaining case where one of the two bands follows the irrep A and the other follows the irrep B. Then, the representation matrix of the $C_6$ rotation  is obtained as
\begin{align}
	D(C_6)=
	\begin{pmatrix}
		1 & 0  \\
		0 & -1 
	\end{pmatrix}.
\end{align}
We shift the origin in $\bm k$ space by $\bm k_0$, so that the focused point becomes $\bm k=0$.

Similar to the previous cases, we begin with a simple example; by using Eq. (\ref{eq:theory_of_invariants}), the $\bm k\cdot\bm p$ Hamiltonian around $\bm k=0$ is written up to the third order in $\bm k$ as
\begin{align}
	\mathcal{H}^{(C_6)}(\bm k)=& a^{(C_6)}_x\sigma_x+a^{(C_6)}_z\sigma_z, \\	
	a^{(C_6)}_x(\bm k)= &c_1k_x(k_x^2-3k_y^2)+c_2k_y(k_y^2-3k_x^2), \label{eq:a_c6x} \\
	a^{(C_6)}_z(\bm k)= &c_3k_z^3+c_4(k_x^2+k_y^2)k_z+c_5(k_x^2+k_y^2) \notag \\
	&+c_6k_z^2+c_7k_z+m, \label{eq:a_c6z}
\end{align}
where $c_i \quad (i=1,2,\cdots ,7)$ and $m$ are real parameters. We set $c_1=1,c_2=1,c_3=0,c_4=0,c_5=1,c_6=-1,c_7=1$ as an example. In $a_x^{(C_6)}(\bm k)$ (Eq. (\ref{eq:a_c6x})) the lowest order terms are of the third order in $\bm k$. On the other hand  the third order terms of $a^{(C_6)}_z(\bm k)$ in Eq. (\ref{eq:a_c6z}) are not essential and they are set to be zero, i.e., $c_3=c_4=0$, for simplicity.
With this choice of parameter values as shown in Fig. \ref{fig:Six-fold-rotation}, nodal lines are reconnected by increasing $m$. Figures \ref{fig:Six-fold-rotation}(a), \ref{fig:Six-fold-rotation}(b), and \ref{fig:Six-fold-rotation}(c) represent nodal lines with $m=-0.75,-0.25,0.25$, respectively. When $m$ is changed from $m=-0.75$ to $m=0.25$, the nodal lines are reconnected at $\bm k=(0,0,\frac{1}{2})$ with $m=-\frac{1}{4}$.

Similar to the case with the $C_4$ symmetry, the topology change of nodal lines cannot be studied within the Morse theory because $\nabla_{\bm k}a_x^{(C_6)}=0$ there. Meanwhile, as is the same with $C_4$ symmetry in Sec. \ref{sec:Result_C4}, we find that the function  $a^{(C_6)}_x$ can always be factorized even in general cases with $C_6$ symmetry, and then the topology change of nodal lines can be fully characterized by the index $\mathcal{N}$ for critical points within the Morse theory. 
Then we find that the evolutions of nodal lines occur on  three surfaces containing the $C_6$ axis, and the events such as reconnection, annihilation, and creation are confined to each surface following $C_6$ symmetry in total as we give detailed explanations in Appendix \ref{app:detail_C6}.

\subsection{Case with the two bands having the same irrep for $C_n$ symmetry ($n=2,4,6$)\label{sec:Cn_with_same_irrep}}

\begin{figure}
	\centering
	\includegraphics[width=\linewidth]{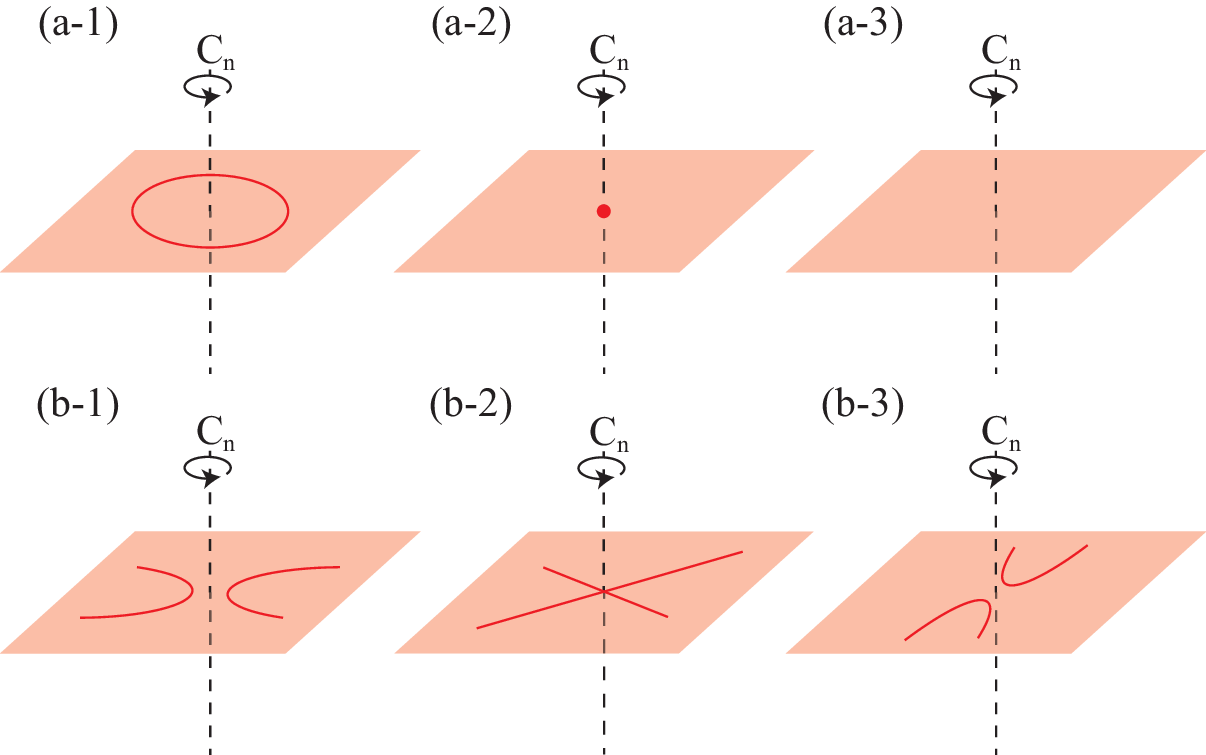}
	\caption{Evolutions of nodal lines with the $C_n$ symmetry ($n=2,4,6$) in the momentum space. The red lines represent nodal lines inside the red plane which is perpendicular to the rotational axis. The annihilation (creation) of nodal lines occurs in (a-1)-(a-3) in the $C_n$ symmetry. The reconnection of nodal lines occurs in (b-1)-(b-3) only in the $C_2$ symmetry, and is prohibited in the $C_4$ and $C_6$ symmetries. }
	\label{fig:illust_Cn}
\end{figure}

We have considered the  nodal lines with the representation matrix composed of two different irreps in $C_n$ symmetry ($n=2,4,6$) so far. In this section, we consider the other case where the two bands have the  same irrep (A or B).  The representation matrix is now written as an identity matrix:
\begin{align}
	D(C_n)=\pm
	\begin{pmatrix}
		1 & 0  \\
		0 & 1 
	\end{pmatrix},
\end{align}
where the plus and minus signs correspond to the A and B irreps, respectively.
Because of Eq. (\ref{eq:theory_of_invariants}), we obtain $a_x(\bm k)=a_x(C_n\bm k)$ and $a_z(\bm k) = a_z(C_n\bm k)$ as symmetry constraints.

The gradients of functions needed for the Morse theory at a point along the $C_n$ rotational axis ($k_z$ axis) are
\begin{align}
	\nabla_{\bm k,m} f(0,0,k_z,m)=(0,0,0,1)^T, \label{eq:nabla_f_Cn} \\
	\nabla_{\bm k,m} a_x(0,0,k_z,m)=(0,0,\cdot,\cdot)^T,  \\
	\nabla_{\bm k,m} a_z(0,0,k_z,m)=(0,0,\cdot,\cdot)^T. \label{eq:nabla_a_Cn_z}
\end{align}
When we focus on the $k_z$ and $m$ components because all the others are zero, the three vectors in Eqs. (\ref{eq:nabla_f_Cn})-(\ref{eq:nabla_a_Cn_z}) reside in a 2D vector space. Therefore, whenever the nodal line cross the $k_z$ axis, this crossing point $(0,0,k_z,m)$ is always a critical point since these three vectors are linearly dependent. Therefore the orthogonal projection $\mathscr{P}$ is given by $\mathscr{P}=\mathrm{diag}(1,1,0,0)$, i.e., a projection to the $k_x$-$k_y$ plane, and the  matrix $\mathscr{M}$ is regarded as a $2\times 2$ matrix within the $k_x$-$k_y$ plane on which $C_n$ symmetry is preserved. In the $C_2$-symmetric case, the two eigenvalues of $\mathscr{M}$ are nonzero and independent in general, and the index $\mathcal{N}$ can have the values 0, 1, or 2. On the other hand, on the $C_4$-symmetric and $C_6$-symmetric cases, the matrix $\mathscr{M}$ has the form $\mathscr{M}=\mathrm{diag}(u,u)$, where $u$ is a real number, due to the above symmetry constraints: $a_x(\bm k)=a_x(C_n\bm k)$ and $a_z(\bm k) = a_z(C_n\bm k)$. Consequently, depending on the sign of $u$, the index $\mathcal{N}$ only gives 0 or 2, corresponding to the creation or the annihilation of nodal lines. In particular, in the $C_4$- and $C_6$- symmetric cases, reconnections of nodal lines never occur on the $C_n$ axis.

In order to see evolutions of nodal lines under the $C_n$ symmetry, we note that $\nabla_{\bm k}a_{x(z)}|_P=(0,0,\cdot)^T$ at the critical point $P$ on the $C_n$ axis defines a normal vector of the surface $a_{x(z)}=0$. Hence, in the vicinity of the critical point $P$, the nodal lines lie along the plane   perpendicular to the rotational axis ($k_z$ axis) as shown in Fig. \ref{fig:illust_Cn}. The nodal lines are illustrated as red lines inside the red plane which is prpendicular to the rotational axis. When $\mathcal{N}=2$,  the annihilation of nodal lines happens, and the nodal lines evolve from Figs. \ref{fig:illust_Cn}(a-1) to \ref{fig:illust_Cn}(a-3) through \ref{fig:illust_Cn}(a-2). The case with $\mathcal{N}=0$ corresponds to the creation, which is a reverse process from Figs. \ref{fig:illust_Cn}(a-3) to \ref{fig:illust_Cn}(a-1). These two processes are allowed in the $C_n$ symmetry ($n=2,4,6$) as we discuss in the previous paragraph. In the case with $\mathcal{N}=1$, corresponding to the reconnection,  the nodal lines evolve from Figs. \ref{fig:illust_Cn}(b-1) to \ref{fig:illust_Cn}(b-3) through \ref{fig:illust_Cn}(b-2). This process is allowed only in systems with $C_2$ symmetry but not with $C_4$ or $C_6$ symmetry. This is naturally understood graphically; by drawing the figures similar to Figs. \ref{fig:illust_Cn}(b-1)-\ref{fig:illust_Cn}(b-3) for $C_n$ ($n=4,6$), one can see that it is improbable for the $n$ nodal lines to meet at the $C_n$ axis. To summarize, when the irreps of the $C_n$ symmetry ($n=2,4,6$) for the conduction and the valence bands are the same, the creation and the annihilation of nodal lines are allowed, but  only the $C_2$ symmetry allows nodal lines to reconnect on the $C_n$ axis.

\subsection{Cases with mirror symmetry\label{sec:Mz}}

\begin{figure}
	\centering
	\includegraphics[width=\linewidth]{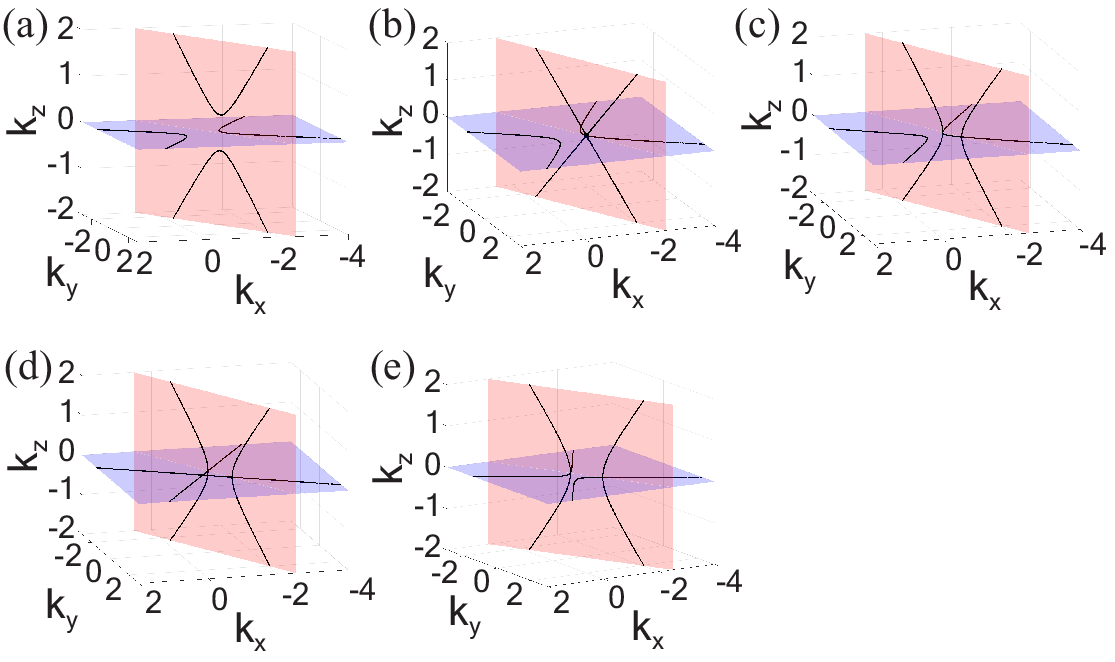}
	\caption{Evolutions of nodal lines with the mirror symmetry in the momentum space, when the two bands have the opposite mirror eigenvalues. Nodal lines are given by the Hamiltonian  $\mathcal{H}^{(M_z)}  (\bm k)$ with $c_1=1,c_2=\frac{1}{2},c_3=1,c_4=1,c_5=-1,c_6=1,c_7=1,c_8=1,c_9=1$. (a), (b), (c), (d) and (e) represent nodal lines with $m=-0.2,-0.05,0.1,0.2,0.3$, respectively. The blue plane is $k_z=0$ plane. The nodal lines are reconnected at $m=-0.05$ and $m=0.2$ on the red plane and the blue plane in (b) and (d), respectively. }
	\label{fig:Mirror}
\end{figure}

In this subsection, we consider nodal lines with \pt\ symmetry and mirror symmetry $M_z$ with respect to the $xy$ plane.
In spinless systems, the eigenvalue of $M_z$ is 1 for the irrep $\mathrm{A}^{'}$ and $-1$ for the irrep $\mathrm{A}^{''}$. 
We discuss topology changes of nodal lines on the mirror-invariant plane. Since the two mirror-invariant planes $k_z=0$ and $k_z=\pi$ can be studied similarly, we focus on the $k_z=0$ plane here.
In this subsection we  consider the case where one of the two bands follows the irrep $\mathrm{A}^{'}$ and the other follows the irrep $\mathrm{A}^{''}$. The other case with the two bands following the same irrep ($\mathrm{A}^{'}$ or $\mathrm{A}^{''}$) will be discussed in Sec. \ref{sec:Mz_with_same_irrep}.  Then, the representation matrix of the mirror operation is
\begin{align}
	D(M_z)=
	\begin{pmatrix}
		1 & 0  \\
		0 & -1 
	\end{pmatrix}.
\end{align}
We focus on a point $\bm k_0$ on the mirror plane, and we shift the origin to the point $\bm k_0$ for simplicity.
By using Eq. (\ref{eq:theory_of_invariants}), we obtain $a_x(k_x,k_y,-k_z)=-a_x(k_x,k_y,k_z)$ and $a_z(k_x,k_y,-k_z)=a_z(k_x,k_y,k_z)$. Thus the $\bm k\cdot\bm p$ Hamiltonian around $\bm k=0$ reflecting $M_z$ symmetry is written up to the second order in $\bm k$ as
\begin{align}
	\mathcal{H}^{(M_z)}(\bm k)=& a^{(M_z)}_x\sigma_x+a^{(M_z)}_z\sigma_z, \label{eq:H_Mz} \\	
	a^{(M_z)}_x(\bm k) = & c_1k_xk_z+c_2k_yk_z+c_3k_z, \label{eq:a_Mz_x} \\
	a^{(M_z)}_z(\bm k) = & c_4k_x^2+c_5k_y^2+c_6k_z^2+c_7k_xk_y \notag \\
	&+c_8k_x+c_9k_y+m, \label{eq:a_Mz_z}
\end{align}
where $c_i \quad (i=1,2,\cdots ,9)$ and $m$ are real parameters. We set $c_1=1,c_2=\frac{1}{2},c_3=1,c_4=1,c_5=-1,c_6=1,c_7=1,c_8=1,c_9=1$ as an example.  Figures \ref{fig:Mirror}(a), \ref{fig:Mirror}(b), \ref{fig:Mirror}(c), \ref{fig:Mirror}(d), and \ref{fig:Mirror}(e) represent nodal lines with $m=-0.2,-0.05,0.1,0.2,0.3$, respectively, and the blue plane is the $k_z=0$ plane. When $m$ is changed from $m=-0.2$ to $m=0.1$, the nodal lines outside of the mirror plane $(k_z=0)$ meet each other at $\bm k=(-\frac{11}{10},\frac{1}{5},0)$ with $m=-0.05$, and they are reconnected (Fig. \ref{fig:Mirror}(b)). At this reconnection, another nodal line on this mirror plane $k_z=0$ also crosses this critical point. Moreover, when $m$ is changed from $m=0.1$ to  $m=0.3$, the nodal lines on the mirror plane $k_z=0$ are reconnected at $\bm k=(-\frac{3}{5},\frac{1}{5},0)$ with $m=0.2$ (Fig. \ref{fig:Mirror}(d)).

\begin{figure}
	\centering
	\includegraphics[width=\linewidth]{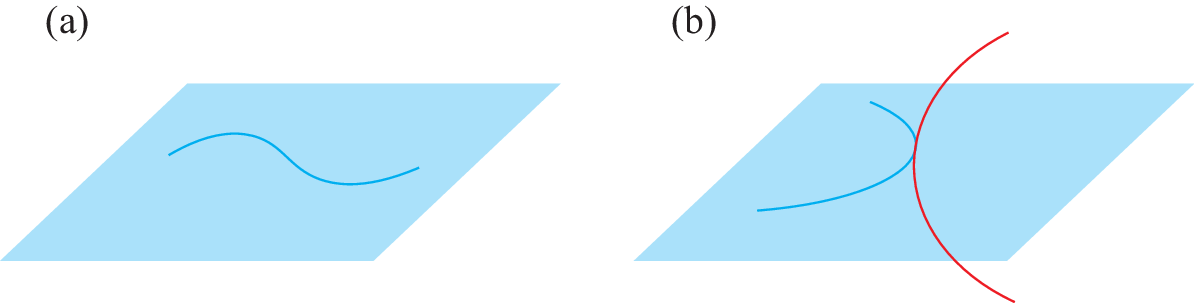}
	\caption{Schematic pictures of existance of nodal lines with the mirror symmetry in the momentum space, when the two bands have the opposite mirror eigenvalues. (a) The blue line represents the ndoal line inside the mirror plane which is illustrated as blue plane. (b) The red line is the nodal line outside the mirror plane.}
	\label{fig:illust_mirror_in_out}
\end{figure}

We now explain the reason why there are two types of reconnections, by noting that $a^{(M_z)}_x(\bm k)$ can be factorized:
\begin{align}
	a^{(M_z)}_x(\bm k) &=k_z\left(k_x+\frac{1}{2}k_y+1\right) \notag \\
	& \equiv k_z \tilde{a}^{(M_z)}_x(\bm k),
\end{align}
where $\tilde{a}^{(M_z)}_x(\bm k)=k_x+\frac{1}{2}k_y+1$.
Therefore, the condition for nodal lines is decomposed into two cases
\begin{enumerate}[label=(\Roman*),align=parleft,leftmargin=50pt,ref=(\Roman*),rightmargin=20pt]
	\item $k_z=0$ and $a^{(M_z)}_z(\bm k)=0$,
	\item $\tilde{a}^{(M_z)}_x(\bm k)=0$ and $a^{(M_z)}_z(\bm k)=0$.	
\end{enumerate}
In  case (I), two nodal lines on the mirror plane ($k_z=0$) meet and reconnect. It occurs at $(\bm k,m)=(-\frac{3}{5},\frac{1}{5},0,\frac{1}{5})$  with index $\mathcal{N}=1$ by the Kamiya theorem corresponding to Fig. \ref{fig:Mirror}(d). 
On the other hand, the case (I\hspace{-1pt}I) describes nodal lines outside of the mirror plane, and at the critical point they meet on the mirror plane $k_z=0$. Therefore, at this point, another nodal line in the case (I)  lying on the mirror plane also goes across the critical point (see Fig. \ref{fig:Mirror}(b)). It occurs at $(\bm k,m)=(-\frac{11}{10},\frac{1}{5},0,-\frac{1}{20})$ with index $\mathcal{N}=1$, corrsponding to the reconnection.

Thus we show that there are two types of nodal-line reconnections from the model (\ref{eq:H_Mz})-(\ref{eq:a_Mz_z}). We can show that also in general systems with mirror symmetry, there are two types, (I) and (I\hspace{-1pt}I), of topology changes of nodal lines from the Morse theory.
Let us assume that $a_x(\bm k)$ and $a_z(\bm k)$  are general analytic functions of $\bm k$ under the $M_z$ symmetry. Even so, $a_x(\bm k)$ is factorized:
\begin{align}
	a_x(\bm k)=k_z\tilde{a}_x(\bm k),
\end{align}
where $\tilde{a}_x(\bm k)$ is an analytic function of $\bm k$ satisfying $\tilde{a}_x(k_x,k_y,-k_z)=\tilde{a}_x(k_x,k_y,k_z)$.
The condition for the nodal line is divided into two cases due to the factorization as follows
\begin{enumerate}[label=(\Roman*),align=parleft,leftmargin=50pt,ref=(\Roman*),rightmargin=20pt]
	\item $k_z=0$ and $a_z(\bm k)=0$,
	\item $\tilde{a}_x(\bm k)=0$ and $a_z(\bm k)=0$.
\end{enumerate}
In cases (I) and (I\hspace{-1pt}I),  the nodal lines appear inside the mirror plane and outside the mirror plane, and they are illustrated as the blue lines on the mirror plane (blue plane) and the red lines outside the mirror plane in Fig. \ref{fig:illust_mirror_in_out}, respectively. In the case (I),  the nodal line exists on the mirror plane as shown in Fig. \ref{fig:illust_mirror_in_out}(a). Meanwhile, in the case (I\hspace{-1pt}I) where the nodal line exists outside the mirror plane, when the nodal line meets the mirror plane  as shown in Fig. \ref{fig:illust_mirror_in_out}(b), the condition of the case (I) is also satisfied there, and it connects with another nodal line on the mirror plane.

\begin{figure}
	\centering
	\includegraphics[width=\linewidth]{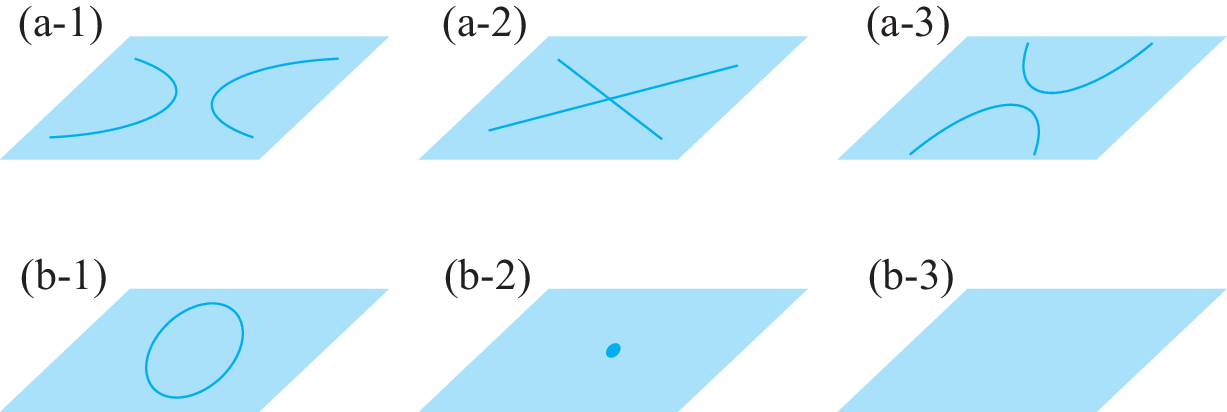}
	\caption{Schematic pictures of evolution of nodal lines inside the mirror plane in the momentum space (case (I)), when the two bands have the opposite mirror eigenvalues. The blue colored lines represent nodal line inside mirror plane, which is illustrated as the blue plane. The reconnection of  nodal lines happens in (a-1)-(a-3), and the annihilation (creation) does in  (b-1)-(b-3).}
	\label{fig:illust_mirror_in}
\end{figure}

We now classify the critical points in terms of the Morse theory. We obtain the gradients of the functions needed for the Morse theory on the mirror plane ($k_z=0$):
\begin{align}
	\nabla_{\bm k,m} f(k_x,k_y,0,m)=(0,0,0,1)^T, \label{eq:nabla_f_Mz} \\
	\nabla_{\bm k,m} \tilde{a}_x(k_x,k_y,0,m)=(\cdot,\cdot,0,\cdot)^T, \label{eq:nabla_a_Mz_x} \\
	\nabla_{\bm k,m} a_z(k_x,k_y,0,m)=(\cdot,\cdot,0,\cdot)^T. \label{eq:nabla_a_Mz_z}
\end{align}
In the case (I), the function $\tilde{a}_x$ is not involved, and Eqs. (\ref{eq:nabla_f_Mz}) and (\ref{eq:nabla_a_Mz_z}) should  be linearly dependent at the critical point, and it holds when $\partial_{k_x}a_z=0$ and $\partial_{k_y}a_z=0$ are satisfied. Therefore,  the number of the conditions (i.e., $\partial_{k_x}a_z=0, \partial_{k_y}a_z=0, a_z=0$)  is three, and it is equal to the number of variables for $(\bm k,m)=(k_x,k_y,0,m)$. This is illustrated as the crossing point on the blue plane in Fig. \ref{fig:Mirror}(d).
On the other hand, in the case (I\hspace{-1pt}I), if there is a point $(k_x,k_y,0,m)$ satisfying
\begin{align}
	\mathrm{Det}
	\begin{pmatrix}
		\partial_{k_x}\tilde{a}_x & \partial_{k_x}a_z \\
		\partial_{k_y}\tilde{a}_x & \partial_{k_y}a_z
	\end{pmatrix}=0, \label{eq:Det_Mz}
\end{align}
that point is regarded as a critical point because  $\nabla_{\bm k,m}f = \alpha_x\nabla_{\bm k,m}\tilde{a}_x + \alpha_z\nabla_{\bm k,m}a_z$ is satisfied for real parameters $\alpha_x$ and $\alpha_z$. The critical point can exist because the number of conditions (i.e., Eq. (\ref{eq:Det_Mz}), $\tilde{a}_x=0$, and $a_z=0$) is three, and it is the same with  the number of variables for $(\bm k,m)=(k_x,k_y,0,m)$. The crossing point on the red plane in Fig. \ref{fig:Mirror}(b) is an example of the above discussion.

\begin{figure}
	\centering
	\includegraphics[width=\linewidth]{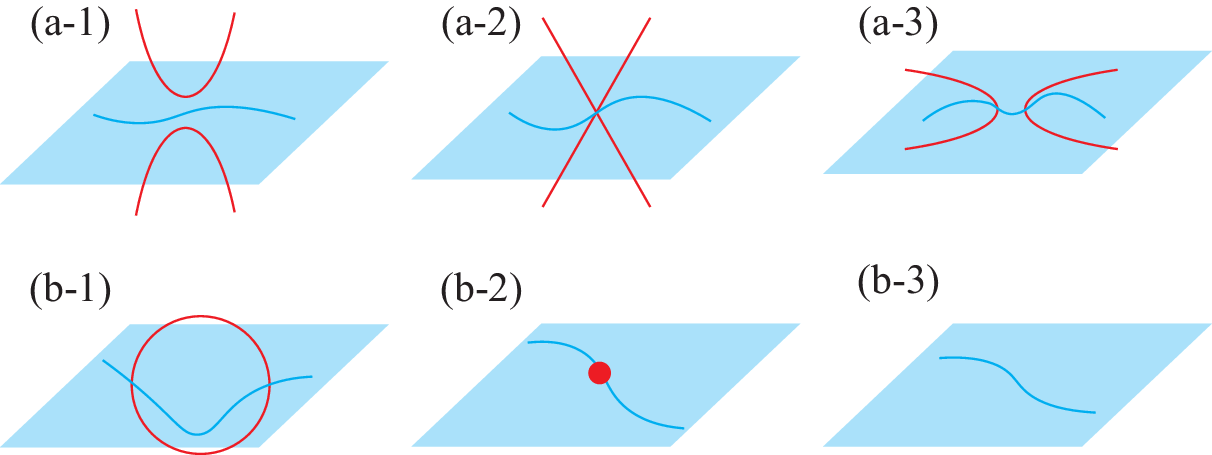}
	\caption{Schematic pictures of evolution of nodal lines outside the mirror plane in the momentum space (case (I\hspace{-1pt}I)), when the two bands have the opposite mirror eigenvalues. The blue lines represent nodal lines inside the mirror plane which is illustrated as blue plane, and the red lines are described as nodal lines outside the mirror plane. When the red colored nodal lines penetrate the mirror plane, they touch with blue colored nodal lines. The reconnection of the nodal lines occurs in (a-1)-(a-3), and  the annihilation (creation) of the nodal lines occurs in (b-1)-(b-3).}
	\label{fig:illust_mirror_out}
\end{figure}

We now discuss topology changes of nodal lines for the cases (I) and (I\hspace{-1pt}I) separately. 
In the case (I), the nodal lines inside the mirror plane evolve as Fig. \ref{fig:illust_mirror_in}. When $\mathcal{N}=1$, the nodal lines are reconnected, and they evolve as Figs. \ref{fig:illust_mirror_in}(a-1)  to \ref{fig:illust_mirror_in}(a-3) through \ref{fig:illust_mirror_in}(a-2). These figures match with the reconnection of nodal lines on the blue plane in Figs. \ref{fig:Mirror}(c), \ref{fig:Mirror}(d), and \ref{fig:Mirror}(e).
When $\mathcal{N}=2$ and the annihilation happens, the nodal lines evolve from Figs. \ref{fig:illust_mirror_in}(b-1) to \ref{fig:illust_mirror_in}(b-3) through \ref{fig:illust_mirror_in}(b-2). Its reverse process is the creation with $\mathcal{N}=0$, where  the nodal ring shrinks to a point on the mirror plane.
Next, in the case (I\hspace{-1pt}I), the nodal lines outside  the mirror plane (red lines in Fig. \ref{fig:illust_mirror_out})  evolve from Figs. \ref{fig:illust_mirror_out}(a-1) to \ref{fig:illust_mirror_out}(a-3) through \ref{fig:illust_mirror_out}(a-2), and vice versa when the reconnection happens, corresponding to $\mathcal{N}=1$. Particularly, in Figs. \ref{fig:illust_mirror_out}(a-2) and \ref{fig:illust_mirror_out}(a-3), the red colored nodal lines cross the blue colored nodal line on the mirror plane because they penetrate mirror plane. These figures explain the reconnection on the red plane in Figs. \ref{fig:Mirror}(a), \ref{fig:Mirror}(b), and \ref{fig:Mirror}(c).
When $\mathcal{N}=2$ and the annihilation happens, the nodal lines evolve from Figs. \ref{fig:illust_mirror_out}(b-1) to \ref{fig:illust_mirror_out}(b-3) through \ref{fig:illust_mirror_out}(b-2). Its reverse process is the creation with $\mathcal{N}=0$, where the nodal ring shrinks to a point on the mirror plane because of the mirror symmetry.

\subsection{Case with two bands having the same irrep for mirror symmetry\label{sec:Mz_with_same_irrep}}

In this subsection, we consider the case where the two bands have the same irrep ($\mathrm{A}^{'}$ or $\mathrm{A}^{''}$) in mirror symmetric systems. The representation matrix is written as
\begin{align}
	D(M_z)=\pm
	\begin{pmatrix}
		1 & 0  \\
		0 & 1 
	\end{pmatrix},
\end{align}
where the plus and minus signs correspond to $\mathrm{A}^{'}$ and $\mathrm{A}^{''}$ irreps, respectively.
and we obtain $a_x(\bm k)=a_x(M_z\bm k)$ and $a_z(\bm k)=a_z(M_z\bm k)$ because of Eq. (\ref{eq:theory_of_invariants}).

The gradients of the functions needed for the Morse theory at a point on the mirror plane ($k_z=0$) are
\begin{align}
	\nabla_{\bm k,m} f(k_x,k_y,0,m)=(0,0,0,1)^T, \label{eq:nabla_f_Mz_same} \\
	\nabla_{\bm k,m} a_x(k_x,k_y,0,m)=(\cdot,\cdot,0,\cdot)^T, \label{eq:nabla_a_Mz_x_same} \\
	\nabla_{\bm k,m} a_z(k_x,k_y,0,m)=(\cdot,\cdot,0,\cdot)^T. \label{eq:nabla_a_Mz_z_same}
\end{align}
If there is a point $(k_x,k_y,0,m)$ satisfying
\begin{align}
	\mathrm{Det}
	\begin{pmatrix}
		\partial_{k_x}a_x & \partial_{k_x}a_z \\
		\partial_{k_y}a_x & \partial_{k_y}a_z
	\end{pmatrix}=0, \label{eq:Det_Mz_same}
\end{align}
the point is a critical point because $\nabla_{\bm k,m}f = \alpha_x\nabla_{\bm k,m}a_x + \alpha_z\nabla_{\bm k,m}a_z$ is satisfied for real parameters $\alpha_x$ and $\alpha_z$. The critical point is determined by three conditions (i.e., Eq (\ref{eq:Det_Mz_same}), $a_x=0$, and $a_z=0$), and these equations can have solutions for three variables (i.e., $k_x$, $k_y$, and $m$). Then, the eigenvalues of the $2\times 2$ matrix $\mathscr{M}$ are nonzero in general, and the critical point is characterized by the index $\mathcal{N}$ into three cases, $\mathcal{N}=0,1,2$, corresponding to creation, reconnection, and annihilation of nodal lines, respectively.

\begin{figure}
	\centering
	\includegraphics[width=\linewidth]{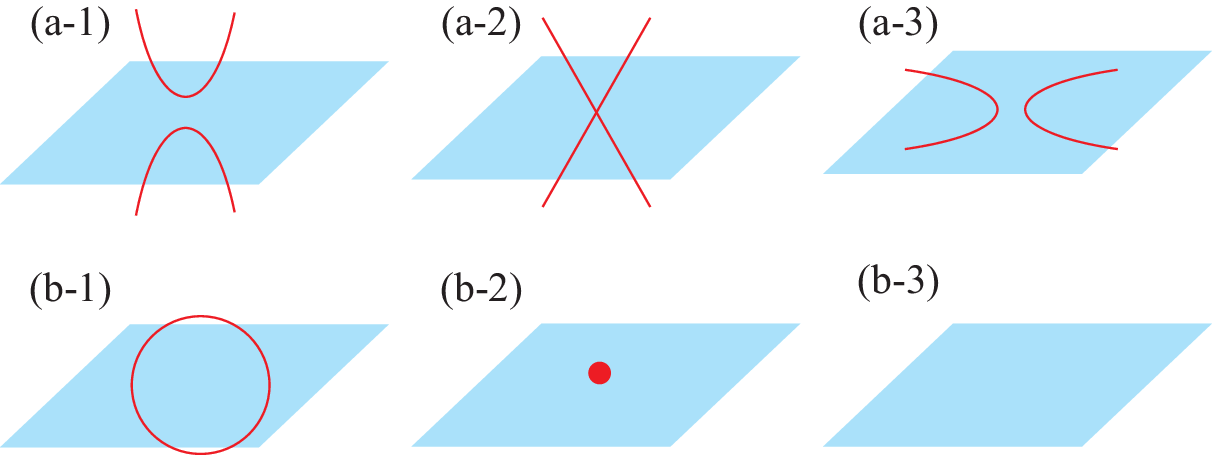}
	\caption{Schematic pictures of evolution of nodal lines outside the mirror plane in the momentum space, when the two bands have the same mirror eigenvalues. The red lines are described as nodal lines outside the mirror plane illustrated as blue plane.  The reconnection of the nodal lines occurs in (a-1)-(a-3), and  the annihilation (creation) of the nodal lines occurs in (b-1)-(b-3).}
	\label{fig:illust_mirror_same}
\end{figure}

In order to see topology changes of nodal lines under the mirror symmetry, we note that the normal vector of the surface $a_{i}=0$ $(i=x,z)$ at any point on the mirror plane is parallel to $\nabla_{\bm k}a_{i}|_P=(\cdot,\cdot,0)^T$. Thus, the tangent vector of the nodal line at any point on the mirror plane is $\bm t||\nabla_{\bm k}a_x\times\nabla_{\bm k}a_z||(0,0,1)$, and it is always perpendicular to the mirror plane as shown in Fig. \ref{fig:illust_mirror_same}. (Note that at a critical point, $\nabla_{\bm k}a_x$ and $\nabla_{\bm k}a_z$ are parallel due to Eq. (\ref{eq:Det_Mz_same}), and the tangent vector $\bm t$ cannot be determined, as is naturally expected.) The red lines represent nodal lines outside the mirror plane which is colored by blue. When $\mathcal{N}=1$, the reconnection of nodal lines occurs, and the nodal lines evolve from Fig. \ref{fig:illust_mirror_same}(a-1) to \ref{fig:illust_mirror_same}(a-3) through \ref{fig:illust_mirror_same}(a-2), and vice versa. When $\mathcal{N}=2$ corrsponding to the annihilation, the nodal line evolves from Figs. \ref{fig:illust_mirror_same}(b-1) to \ref{fig:illust_mirror_same}(b-3) through \ref{fig:illust_mirror_same}(b-2). In the case with $\mathcal{N}=0$, the nodal line is created as a reverse process of the annihilation.

\section{Summary\label{sec:summary}}
In this paper, we reveal that there are three types of evolutions of nodal lines with PT symmetry such as the creation, reconnection, and annihilation. Such topology changes of nodal lines are understood by the local maximum, the saddle point, or the local minimum of the function $f(\bm k,m)=m$ in the 4D ($\bm k,m$) space. These critical points are characterized by the index $\mathcal{N}=0,1,2$ corresponding to the creation, reconnection, and annihilation in the Morse theory, and we give examples for the transition of nodal lines. As a result, we show that a phase transition between the nodal lines and the nodal link cannot occur directly, but via several reconnections of nodal lines, if no additional crystallographic symmetries are assumed. 

Moreover we extend our theory to the case with the rotational symmetries and the mirror symmetry, and disclose the possible topology changes of nodal lines. 
When the system has $C_n$ symmetry $(n=2,4,6)$, the events of topology changes of nodal lines occur inside the plane containing (perpendicular to) the $C_n$ axis with the different (the same) irreps for the two band. We exclude the $C_3$ symmetric case because the nodal line always lies along the rotational axis. When the two bands have different irreps with $C_n$ symmetry $(n=2,4,6)$, all events for the topology changes of nodal lines (i.e., creation, reconnection, and annihilation) are possible. Meanwhile, with the same irreps for the two bands, only creation and annihilation are allowed in $n=4,6$ because of the symmetry constraints, whereas  all events are allowed on $n=2$.
When the system has the mirror symmetry with two bands having the different irreps, the topology changes of nodal lines classified into those for nodal lines inside and outside of the mirror plane, corresponding to cases (I) and (I\hspace{-1pt}I) in Sec. \ref{sec:Mz} respectively. On the other hand, if the two bands have the same irreps, the nodal lines reside only outside the mirror plane, and can experience all possible topology changes.

Thus, we have shown that the topology changes of nodal lines are determined by the point-group symmetry of the $\bm k$ point at which the topology change occurs. In this paper, we limit ourselves to the simplest point groups such as rotation or mirror reflection only. An extension to other point groups is beyond of this paper, and is left as a future work.
In this paper, we limit ourselves to the nodal lines composed of one conduction band and one valence band, and we describe the nodal lines in terms of the two-band model. It should be interesting if we extend our analysis to the cases with the larger number of bands, which will include various intriguing cases such as the triple points \cite{PhysRevB.103.L121101} and double band inversion \cite{PhysRevLett.121.106403}. Nonetheless, the analysis for the larger number of bands will be largely different and much more complicated than the two-band cases. Therefore the extension to multiband systems is left for future work.

\begin{acknowledgments}
This work is supported by JSPS KAKENHI Grant Numbers JP20J23473, JP22K18687, and JP22H00108.
\end{acknowledgments}

\appendix
\section{Kamiya theorem \label{app:Kamiya_theorem}}

The nodal lines evolve by changing the parameter $m$, and all possible nodal lines configure the manifold $M$ in $(\bm k,m)$ space.  The topology changes of nodal lines occur across the critical points on the manifold $M$. To identify the local structure of the manifold $M$ around the critical points, we adopt the Kamiya theorem in Secs. \ref{sec:evolution_no_sym} and \ref{sec:evolution_with_sym}. In this Appendix, we explain the Kamiya theorem.

The Kamiya theorem is as follows \cite{oai:soar-ir.repo.nii.ac.jp:00012243}. Let  $M$ be a differenctiable function described as
\begin{align}
	M=\{ Q\in\mathbb{R}^n| f_1(Q)=\cdots =f_r(Q)=0 \},
\end{align}
where $f_1,\cdots f_r:\mathbb{R}^n\to\mathbb{R}$ are differentiable functions, and $\nabla f_1(Q),\cdots ,\nabla f_r(Q)$ be linearly independent.
If and only if  for a given function $f: \mathbb{R}^n\to \mathbb{R}$, the following equation holds
\begin{align}
	\nabla f(Q_0)=a_1\nabla f_1(Q_0)+\cdots +a_r\nabla f_r(Q_0) \quad  a_i\in \mathbb{R},
\end{align}
$Q_0\in M$ is a critical point of the function $\bar{f}=f|_M:M\to \mathbb{R}$.
Then, let $Q_0\in M$ be a critical point of $\bar{f}$, and $\mathscr{P}$ be a orthogonal projection from $\mathbb{R}^n$ to a tangent vector space $T_{Q_0}(M)$ at $Q_0$ i.e., $\mathscr{P}:\mathbb{R}^n\to T_{Q_0}(M)$.
If and only if  a rank of a matrix
\begin{align}
	\mathscr{M}=\mathscr{P}(\mathscr{H}(f)|_{Q_0}-\sum_{i=1}^{r}a_i\mathscr{H}(f_i)|_{Q_0})\mathscr{P},
\end{align}
where $\mathscr{H}$ is a Hessian matrix, is $n-r$, i.e., $\text{rank}\  \mathscr{M}=n-r$, the critical point $Q_0$ is nondegenerate. Moreover, an index of $f$ at $Q_0$ is equal to the number of negative eigenvalues of matrix $\mathscr{M}$.
In the main text, we take $n=4$, $r=2$, $f_1=a_x$, and $f_2=a_z$.

\section{Morse lemma \label{app:Morse_theory}}
The Morse lemma \cite{Audin_2014,Nicolaescu_2011} used in Secs. \ref{sec:evolution_no_sym} and \ref{sec:evolution_with_sym} tells us the local shape of the function $f$ near its critical point.
Let $M$  be a differentiable function and  $f: M \to \mathbb{R}$ be a diferentiable function. When a point $Q\in M$ is a nondegenerate critical point of $f$, the function $f$ can be expressed in terms of a local cordinate ($U;x_1,\cdots ,x_n$) around $Q$:
\begin{align}
	&x_1(Q)=\cdots =x_n(Q)=0, \\
	&f=f(Q)-x_1^2-\cdots -x_r^2+x_{r+1}^2+\cdots +x_n^2,
\end{align}
where $r$ is a index of $f$ on $Q$.

\section{Detailed calculation for the index of the annihilation case}

We show the detailed calculations of the model for the annihilation.
As the gradients of $a^{\mathrm{(C)}}_j(\bm k,m)$ for $j=x,z$ at $(\bm k,m)=(0,0)$ are
\begin{align}
\nabla_{\bm k,m}a^{\mathrm{(C)}}_x(0,0)  =
\begin{pmatrix}
1\\
0\\
0\\
2
\end{pmatrix}, \ 
\nabla_{\bm k,m}a^{\mathrm{(C)}}_z(0,0)  =
\begin{pmatrix}
1\\
0\\
0\\
1
\end{pmatrix}, \label{eq:normal_vec_H_C}
\end{align}
we get the following relationship:
\begin{align}
\nabla_{\bm k,m}f(0, 0)=\nabla_{\bm k,m}a^{\mathrm{(C)}}_x(0,0)-\nabla_{\bm k,m}a^{\mathrm{(C)}}_z(0,0).
\end{align}
This equation implies that $Q(0,0)$ is a critical point by the Kamiya theorem.
The vectors in Eq. (\ref{eq:normal_vec_H_C}) span a 2D vector space with basis vectors 
\begin{align}
\bm b^{\mathrm{(C)}}_1  =
\begin{pmatrix}
1\\
0\\
0\\
0
\end{pmatrix}, \ 
\bm b^{\mathrm{(C)}}_2  =
\begin{pmatrix}
0\\
0\\
0\\
1
\end{pmatrix},
\end{align}
and an orthogonal projection to a tangent vector space at the critical point $Q(0,0)$ is written as
\begin{align}
\mathscr{P}^{\mathrm{(C)}}  & = I_4-\bm b^{\mathrm{(C)}}_{ 1}\bm b^{\mathrm{(C)}T}_{ 1}-\bm b^{\mathrm{(C)}}_{ 2}\bm b^{\mathrm{(C)}T}_{ 2}  \\
&= \begin{pmatrix}
0 &  &  & \\
& 1&  & \\
&  & 1& \\
&  &  & 0
\end{pmatrix}.
\end{align}
Moreover  Hessian matrices of $a^{\mathrm{(C)}}_{ j}(\bm k, m)$ and $f(\bm k,m)$ at the critical point $Q$ are obtained as
\begin{align}
&\mathscr{H}(a^{\mathrm{(C)}}_{ x})|_Q=
\begin{pmatrix}
1 &  &  &  \\
& 2&  &  \\
&  & 2& \\
&  &  & 1
\end{pmatrix}, \\
&\mathscr{H}(a^{\mathrm{(C)}}_{z})|_Q=
\begin{pmatrix}
-1&  &  &  \\
& 1&  &  \\
&  & -1&  \\
&  &  &-1
\end{pmatrix}, \\
&\mathscr{H}(f)|_Q=0.
\end{align}
A matrix $\mathscr{M}^{\mathrm{(C)}}$ written as 
\begin{align}
\mathscr{M}^{\mathrm{(C)}} & = \mathscr{P}^{\mathrm{(C)}} (\mathscr{H}(f)|_Q-\mathscr{H}(a^{\mathrm{(C)}}_{x})|_Q+\mathscr{H}(a^{\mathrm{(C)}}_{z})|_Q)\mathscr{P}^{\mathrm{(C)}} \notag \\
& = 
\begin{pmatrix}
0&  &  &  \\
&-1&  &  \\
&  & -3&  \\
&  &  &0
\end{pmatrix}
\end{align}
indicates that the critical point $Q$ is nondegenerate  by the Kamiya theorem  because the rank of $\mathscr{M}^{\mathrm{(C)}}$ is 2. Furthermore, as the number of the negative eigenvalues is equal to an index of $f$ at the critical point $Q$, the index $\mathcal{N}$ of $f(Q)$ is 2.

\section{Further explanations for the nodal link\label{app:further_explanation_nodal_link}}

\begin{figure}
	\centering
	\includegraphics[width=\linewidth]{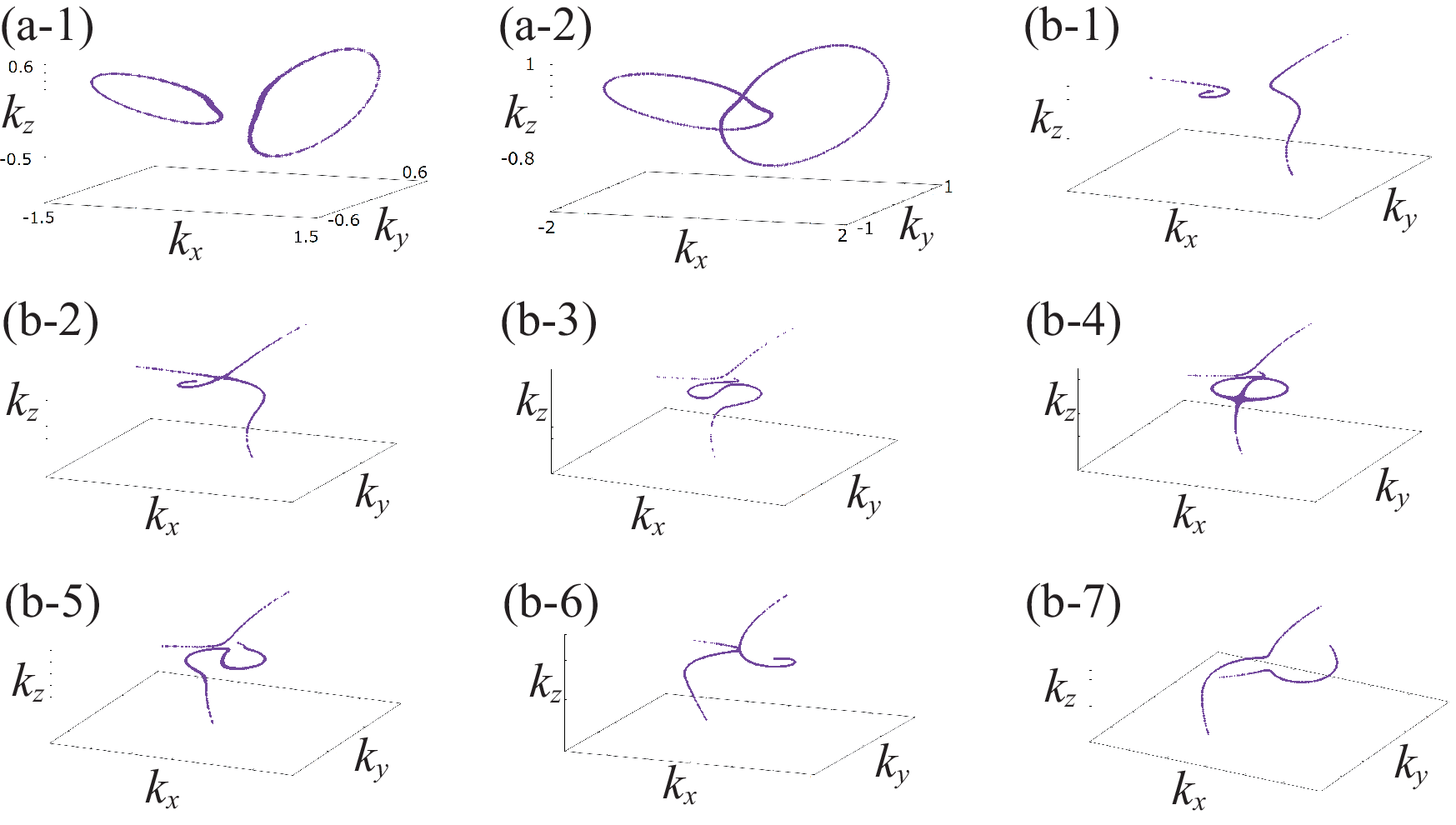}
	\caption{Nodal lines in the Hamiltonian $\mathcal{H}^{(\mathrm{A})}$ with an additional term. There are two nodal lines with $m=3.2$ in (a-1) and ther is the nodal link with $m=2.8$ in (a-2). The intermediate states between the nodal lines and the nodal link are illustrated in (b-1)-(b-7) as enlarged views. The nodal lines with $m=3.07,\ 3.05,\ 3,\ 2.995,\ 2.99,\ 2.95,\ \text{and}\ 2.93$ are illustrated in (b-1)-(b-7) in this order. From the two nodal lines to a nodal link, reconnections occur three times in (b-2), (b-4) and (b-6).}
	\label{fig:nodal_link-nodal_line}
\end{figure}

As we discussed in Sec. \ref{sec:previous-nodal-link}, the model (\ref{eq:H_A})-(\ref{eq:a_Az}) gives a direct phase transition between the nodal lines and the nodal link, but in terms of the Morse theory it is an artifact of the special choice of the model. Namely,   $\nabla_{\bm k}a^{(A)}_x=\nabla_{\bm k}a^{(A)}_z=0$ holds at the transition where the two nodal lines touch, but these equations cannot hold simultaneously in general, and they hold just by accident. To remove this artifact, we add some terms to the Hamiltonian: $a^{(A)}_x\to a'^{(A)}_x=a^{(A)}_x+\alpha\sin k_x$ and $a^{(A)}_z\to a'^{(A)}_z=a^{(A)}_z+\alpha\sin k_z$, where $\alpha$ is a real parameter.
This model has the nodal lines with $m=3.2$ and the nodal link with $m=2.8$ in Figs. \ref{fig:nodal_link-nodal_line}(a-1) and \ref{fig:nodal_link-nodal_line}(a-2) like the original model $\mathcal{H}^{(\mathrm{A})}$. However, the intermediate states are different from the original model as shown in Figs. \ref{fig:nodal_link-nodal_line}(b-1)-\ref{fig:nodal_link-nodal_line}(b-7), where the nodal lines with $m=3.07, 3.05, 3, 2.995, 2.99, 2.95, 2.93$ are illustrated. The nodal lines are reconnected three times in Figs. \ref{fig:nodal_link-nodal_line}(b-2), \ref{fig:nodal_link-nodal_line}(b-4), and \ref{fig:nodal_link-nodal_line}(b-6). Therefore, this model requires the reconnections of nodal lines three times to get the nodal link from two nodal lines. 

\begin{figure}
	\centering
	\includegraphics[width=\linewidth]{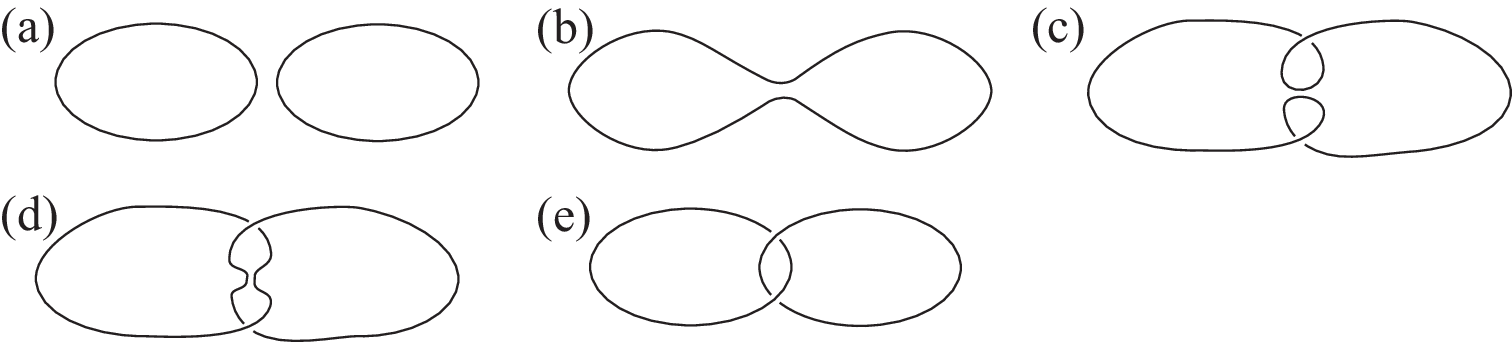}
	\caption{Schematic pictures of evolution  from the nodal lines in (a) to the nodal link (e). The reconnections happen twice as from (a) to (b) and from (c) to (d).}
	\label{fig:illust_nodal_link-nodal_line}
\end{figure}

Thus the phase transition from two nodal lines to a nodal link cannot occur directly but via several reconnections. We show its simplest pattern schematically in Fig. \ref{fig:illust_nodal_link-nodal_line} from (a) to (e) as an example. In this example, the nodal lines are reconnected twice between (a) and (b), and (c) and (d), and then   the nodal link is finally obtained.

\section{Factorization of $a_x$}
Here, we show that $a_x$ is always factorized when the two bands have different irreps for $C_4$ and $C_6$ symmetries, as briefly explained in Secs.  \ref{sec:Result_C4} and \ref{sec:IVA4}.
On the premise that $a_x(\bm k)$ is an analytic function of $k_x$ and $k_y$, one can write
\begin{align}
	a_x(\bm k)=\sum_{m=0}^{\infty}\sum_{n=0}^{\infty}c_{mn}k_+^mk_-^n,
\end{align}
where $c_{mn}$ is a complex analytic function of $k_z$, and $k_{\pm}=k_x\pm ik_y$. For the following calculations, we note that $c_{nm}=\bar{c}_{mn}$ since $a_x(\bm k)$ is real.

\subsection{$C_4$ symmetry \label{app:C4}}
When the system has $C_4$ symmetry, we obtain $a_x(k_+,k_-)=-a(ik_+,-ik_-)$ because of Eq. (\ref{eq:theory_of_invariants}), and  the summation in $a_x(\bm k)$ is rewritten as
\begin{align}
	a_x(\bm k)&=\sum_{\substack{m\geq 0 \\ n\geq 0 \\ m-n\equiv 2\ (\mathrm{mod}\ 4)}}c_{mn}k_+^mk_-^n, \notag \\
	&=\left( \sum_{\substack{m>n\geq 0 \\ m-n\equiv 2\ (\mathrm{mod}\ 4)}} + \sum_{\substack{n>m\geq 0 \\ m-n\equiv 2\ (\mathrm{mod}\ 4)}} \right)c_{mn}k_+^mk_-^n, \notag \\
	&=c_{20}k_+^2\sum_{\substack{p=0 \\ q=0}}^{\infty}d_{pq}(k_+k_-)^pk_+^{4q} + c_{02}k_-^2\sum_{\substack{p=0 \\ q=0}}^{\infty}\bar{d}_{pq}(k_+k_-)^pk_-^{4q},
\end{align}
where $d_{pq}=c_{p+4q+2,p}/c_{20}$.  
Because $d_{00}=1$, a square root  of the sum $\sum_{\substack{p=0 \\ q=0}}^{\infty}d_{pq}(k_+k_-)^pk_+^{4q}$ is analytic in $k_+$ and $k_-$, and we can write $g^2(\bm k)=\sum_{\substack{p=0 \\ q=0}}^{\infty}d_{pq}(k_+k_-)^pk_+^{4q}$, where $g(\bm k)$ is analytic and $g(\bm k=0)=1$. When we introduce $i\alpha^2=c_{20}$, where $\alpha$ is complex constant, $a_x(\bm k)$ is explicitly factorized as follows:
\begin{align}
	a_x(\bm k)&=i\alpha^2 k_+^2g^2(\bm k)-i\bar{\alpha}^2k_-^2\bar{g}^2(\bm k), \notag \\
	&=i(\alpha  k_+g(\bm k)-\bar{\alpha}k_-\bar{g}(\bm k))(\alpha  k_+g(\bm k)+\bar{\alpha}k_-\bar{g}(\bm k)).
\end{align}
Because  $g(\bm k)$ is written as $g(\bm k)=\sum_{\substack{p=0 \\ q=0}}^{\infty}f_{pq}(k_+k_-)^pk_+^{4q}$, where $f_{pq}$ is a complex constant, we get $g(\bm k)=g(C_4\bm k)$. 

\subsection{$C_6$ symmetry}
When the system has $C_6$ symmetry, we obtain $a_x(k_+,k_-)=-a_x(\mathrm{e}^\frac{\pi i}{3}k_+,\mathrm{e}^{-\frac{\pi i}{3}}k_-)$ due to Eq. (\ref{eq:theory_of_invariants}). Therefore, as well as the $C_4$-symmetric case, $a_x(\bm k)$ is factorized:
\begin{align}
	a_x(\bm k)=& i\alpha^3 k_+^3g^3(\bm k)-i\bar{\alpha}^3k_-^3\bar{g}^3(\bm k), \notag \\
	=& i(\alpha k_+g(\bm k)-\bar{\alpha} k_-\bar{g}(\bm k)) \notag \\
	&(\alpha k_+\mathrm{e}^{\frac{2}{3}\pi i}g(\bm k)-\bar{\alpha} k_-\mathrm{e}^{-\frac{2}{3}\pi i}\bar{g}(\bm k)) \notag \\
	&(\alpha k_+\mathrm{e}^{\frac{4}{3}\pi i}g(\bm k)-\bar{\alpha} k_-\mathrm{e}^{-\frac{4}{3}\pi i}\bar{g}(\bm k)),
\end{align}
where $\alpha$ is a complex constant, $g(\bm k)$ is an analytic function of $k_x$ and $k_y$ with $g(\bm k=0)=1$, and $g(\bm k)=g(C_6\bm k)$.

\section{Detail explanation under $C_6$ symmetry \label{app:detail_C6}}
In this Appendix, we show the details of the factorization of $a_x^{(C_6)}$ under $C_6$ symmetry, similar to the cases under $C_4$ symmetry.
In order to study the topology change of nodal lines within the Morse theory,  we will show that $a^{(C_6)}_x$ can be factorized because of the $C_6$ symmetry. For example, when $a_x^{(C_6)}$ is given by Eq. (\ref{eq:a_c6z}) with $c_1=1$ and $c_2=1$, it is factorized as
\begin{align}
	a^{(C_6)}_x(\bm k) & =(k_x+k_y)((2-\sqrt{3})k_x-k_y)((2+\sqrt{3})k_x-k_y) \notag \\
	& \equiv a^{(C_6)(\mathrm{I})}_x(\bm k)a^{(C_6)(\mathrm{I}\hspace{-1pt}\mathrm{I})}_x(\bm k)a^{(C_6)(\mathrm{I}\hspace{-1pt}\mathrm{I}\hspace{-1pt}\mathrm{I})}_x(\bm k),
\end{align}
where $a^{(C_6)(\mathrm{I})}_x(\bm k)=k_x+k_y$, $a^{(C_6)(\mathrm{I}\hspace{-1pt}\mathrm{I})}_x(\bm k)=(2-\sqrt{3})k_x-k_y$,  and $a^{(C_6)(\mathrm{I}\hspace{-1pt}\mathrm{I}\hspace{-1pt}\mathrm{I})}_x(\bm k)=(2+\sqrt{3})k_x-k_y$.
Therefore, the condition for nodal lines is decomposed into three cases:
\begin{enumerate}[label=(\Roman*),align=parleft,leftmargin=50pt,ref=(\Roman*),rightmargin=20pt]
	\item $a^{(C_6)(\mathrm{I})}_x(\bm k)=0$ and $a^{(C_6)}_z(\bm k)=0$ ,
	\item $a^{(C_6)(\mathrm{I}\hspace{-1pt}\mathrm{I})}_x(\bm k)=0$ and $a^{(C_6)}_z(\bm k)=0$,
	\item $a^{(C_6)(\mathrm{I}\hspace{-1pt}\mathrm{I}\hspace{-1pt}\mathrm{I})}_x(\bm k)=0$ and $a^{(C_6)}_z(\bm k)=0$.
\end{enumerate}
In all three cases (I), (I\hspace{-1pt}I), and (I\hspace{-1pt}I\hspace{-1pt}I), one can check that the point $(\bm k,m)=(0,0,\frac{1}{2},-\frac{1}{4})$ is a critical point with the index $\mathcal{N}=1$ by the Kamiya theorem. Then the reconnections of nodal lines on the three planes $a^{(C_6)(\mathrm{I})}_x(\bm k)=k_x+k_y=0$, $a^{(C_6)(\mathrm{I}\hspace{-1pt}\mathrm{I})}_x(\bm k)=(2-\sqrt{3})k_x-k_y=0$, and $a^{(C_6)(\mathrm{I}\hspace{-1pt}\mathrm{I}\hspace{-1pt}\mathrm{I})}_x(\bm k)=(2+\sqrt{3})k_x-k_y=0$  occur simultaneously by the Morse lemma in a $C_6$-symmetric manner.

Apart from the above example, even when $a_x(\bm k)$ and $a_z(\bm k)$ are general analytic functions around the critical point $(0,0,k_z^{(0)})$ in $\bm k$ with the higher order terms in $\bm k$, we can show that $a_x(\bm k)$ is factorized:
\begin{align}
	a_x(\bm k)=&i (\alpha k_{+}g(\bm k)-\bar\alpha k_{-}\bar{g}(\bm k))  \notag \\
	&(\alpha k_+\mathrm{e}^{\frac{2}{3}\pi i}g(\bm k)-\bar{\alpha} k_-\mathrm{e}^{-\frac{2}{3}\pi i}\bar{g}(\bm k)) \notag \\
	&(\alpha k_+\mathrm{e}^{\frac{4}{3}\pi i}g(\bm k)-\bar{\alpha} k_-\mathrm{e}^{-\frac{4}{3}\pi i}\bar{g}(\bm k)), \notag \\
	\equiv& i  a^{(\mathrm{I})}_x(\bm k)a^{(\mathrm{I}\hspace{-1pt}\mathrm{I})}_x(\bm k)a^{(\mathrm{I}\hspace{-1pt}\mathrm{I}\hspace{-1pt}\mathrm{I})}_x(\bm k),
\end{align}
where $a^{(\mathrm{I})}_x(\bm k)=\alpha k_{+}g(\bm k)-\bar\alpha k_{-}\bar{g}(\bm k)$, $a^{(\mathrm{I}\hspace{-1pt}\mathrm{I})}_x(\bm k)=\alpha k_+\mathrm{e}^{\frac{2}{3}\pi i}g(\bm k)-\bar{\alpha} k_-\mathrm{e}^{-\frac{2}{3}\pi i}\bar{g}(\bm k)$,  $a^{(\mathrm{I}\hspace{-1pt}\mathrm{I}\hspace{-1pt}\mathrm{I})}_x(\bm k)=\alpha k_+\mathrm{e}^{\frac{4}{3}\pi i}g(\bm k)-\bar{\alpha} k_-\mathrm{e}^{-\frac{4}{3}\pi i}\bar{g}(\bm k)$ is an analytic function of $k_\pm$ and $k_z$ satisfying $g(0,0,k_z^{(0)})=1, g(\bm k)=g(C_6\bm k)$, and $\alpha$ is a complex constant.
As well as the $C_4$-symmetric case, the condition for nodal lines is decomposed into three cases, and $a^{(i)}_x(\bm k)=0\ (i=\mathrm{I},\mathrm{I}\hspace{-1pt}\mathrm{I},\mathrm{I}\hspace{-1pt}\mathrm{I}\hspace{-1pt}\mathrm{I})$ represent three $C_2$-symmetric surfaces, which are $C_6$-symmetric in total. Therefore, the evolutions of nodal lines occur on  three surfaces containing the $C_6$ axis, and the events such as reconnection, annihilation, and creation are confined to each surface, and they follow $C_6$ symmetry in total.
The topology change can be illustrated easily. It is similar to Fig. \ref{fig:illust_four-roto}, only with the change from $C_4$ to $C_6$, and so is omitted here.


%

\end{document}